\documentclass[12pt,showpacs,amssymb,aps,nofootinbib,floatfix]{revtex4-1}
\usepackage{epsfig}
\usepackage{xcolor}
\usepackage{amsmath}
\usepackage{slashed}
\usepackage[T1]{fontenc}
\usepackage[utf8]{inputenc}
\usepackage[english]{babel}
\newcommand{\bra}[1]{\left< #1 \right| }
\newcommand{\order}[1]{ \mathcal{O} \left( #1 \right) }
\newcommand{\ket}[1]{\left| #1 \right> }

\newcommand{\eqcomma}{\phantom{AA},\phantom{AA}}

\begin{document}
\title{Meson spin alignment and baryon polarization from coalescence with spin-vorticity non-equilibrium}
\author{Kayman J. Gon\c{c}alves$^{1}$, Giorgio Torrieri$^{1}$, Radoslaw Ryblewski$^{2}$}
\affiliation{$\phantom{A}^{1}$IFGW, Unicamp}
\affiliation{$\phantom{A}^{2}$Institute of Nuclear Physics, Polish Academy of Sciences, PL-31-342 Krakow, Poland}
\begin{abstract} 
We estimate both the polarization of spin $1/2$ baryons and the spin density matrix coefficients of spin $1$ mesons using a  thermal model that incorporates vorticity and polarization to describe quarks, along with a coalescence formalism where spin and vorticity are not in equilibrium.   We find that while our model is not predictive due to its considerable number of free parameters, it has the potential to explain several seemingly puzzling features of experimental spin measurements, such as the absence of baryon polarization alongside significant vector meson spin alignment.   We suggest the measuring meson spin off-diagonal matrix elements, and examining the dependence of polarization observables on azimuthal angles, as methods to falsify this model and gain insights into the freezeout details of baryon and meson spin structure.
\end{abstract}
\maketitle
\section{Introduction and Motivation}
The measurements of baryon polarization \cite{lambda} and vector meson spin alignment \cite{vector,charmonium} have haralded a fertile era of experimental and theoretical study of relativistic hydrodynamics with vorticity \cite{karpenko,review,spinshear}.

On one hand, it is encouraging that hydrodynamics together with a Cooper-Frye formula extended to spinors \cite{zanna} can describe global polarization data \cite{review} based on freeze-out from a vortical fluid.  On the other hand, some phenomenological issues remain: 
\begin{itemize}
\item {\em Local} polarization appears to have the wrong sign \cite{review} \item Vector meson spin alignment is not described by theoretical models \cite{vector}, which has been interpreted as suggesting new physics such as classical fields \cite{classic0,classic1,classic2,classic3} (suggesting the $\phi$ is coherent) and spin distillation \cite{kayman2} (which would mean the $\phi$ is more akin to a quarkonium state).
\end{itemize}
In addition, while \cite{karpenko,review} treat polarisation and vorticity as quantities related by an equilibrium relation (as \cite{tinti,friman}), it is clear that causality forbids spin vorticity from being in full equilibrium \cite{causal,montediss}. There is now a consensus among theorists that a hydrodynamics framework incorporating both vorticity and spin is necessary.   Several versions of such hydrodynamics have been proposed \cite{relax,jeon,rischke,hongo,friman,florkowski} making it imperative to establish a connection with phenomenology, specifically regarding the issues discussed in the previous paragraph.

As previously argued \cite{kayman1,kayman2} vector mesons are a promising probe in this regard, as their $3X3$ density matrix contains quantifiable traces of quantum coherence that can be directly connected to experimental observables \cite{seyboth,wang}. 

Specifically, the spin $1/2$ baryon angular distribution in its rest frame is governed by a single coefficient\footnote{$W(...)$ refers to the probability of the daughter to decay. As we explain later it needs to be convoluted with the parent distribution.}  where $\phi^*$ is the daughter particle emission angle measured with respect to parent particle spin axis in its rest frame.
\begin{equation}
\label{spinlambda}
W(\theta^*,\phi^*) \propto 1+ \chi_{1/2} \cos{\phi^*} \eqcomma \chi_{1/2} = \rho_{1/2,1/2}
\end{equation}
On the other hand, for vector mesons,
provided we can reliably determine a common axis of the $\theta^*$ angle perpendicular to that of $\phi^*$ for all events (in other words if the rapidity distribution of the parent vector meson is understood), one can obtain the $\theta^*, \phi^*$ distribution in the rest frame of the parent particle and relate it to the density matrix
\begin{equation}
W(\theta^*,\phi^*) \propto
\label{dis}  \cos^{2}\phi^*\rho_{00}+\sin^{2}\phi^* \left(  \frac{1-\rho_{00}}{2}+  r_{1,-1} \cos(2\theta^*)+\alpha_{1,-1} \sin(2\theta^*) \right)
 \end{equation}
 \[
    + \sin(2 \phi^*) \left( r_{10} \cos\theta^*+\alpha_{10} \sin\theta^*\right)
 \]
The coefficients of the trigonometric expressions in the above formula are related to the density matrix elements of the meson ensemble through a normalization factor $3/(4\pi)$
\begin{equation}
  \begin{array}{ccc}
\mathrm{Variable} &    \mathrm{Element}& \mathrm{coefficient}\times \frac{3}{4\pi}\\
 \rho^M_{00} &    \rho^M_{00} & \cos^2\phi^*\\
 \frac{1-\rho^M_{00}}{2} &   \frac{\rho^M_{11}+\rho^M_{-1-1}}{2} & \sin^2 \phi^*\\
    r_{10} & \text{Re}[\rho^M_{-10}-\rho^M_{10}] & \sin(2\phi^*)\cos(\theta^*)\\
   \alpha_{10} &   \text{Im}[-\rho^M_{-10}+\rho^M_{10}] & \sin(2\phi^*)\sin(\theta^*)\\
   r_{1,-1} &     \text{Re}[\rho^M_{1,-1}] & \sin^2 \phi^* \cos (2\theta^*)\\
   \alpha_{1,-1} &     \text{Im}[\rho^M_{1,-1}] & \sin^2 \phi^* \sin(2\theta^*)
  \end{array}
  \label{tablecoeff}
\end{equation}
Notably, it has long been observed \cite{kayman1,oliva,coalphi,xia} that these off-diagonal matrix elements can provide insights into the hadronization of vector mesons.   

In this work we develop this idea by combining a previously used blast wave model augmented with vorticity \cite{FLO} with the coalescence ansatz of \cite{kayman1} to attempt to describe both vector meson spin alignment, including non-diagonal terms, and baryon polarization.
We note that our aim is not quantitative precision as there are considerable phenomenological uncertainities.  This is because
\begin{itemize}
\item The model in \cite{FLO} has not been tuned for the coalescence dynamics of \cite{kayman1}
\item The model in \cite{kayman1} relies on parameters that we do not know how to estimate from first principles, such as the spin projection of the coalescence Wigner function and the projector $P_L$ which regulates the angular momentum transfer from vorticity to spin. This parameter could in principle be calculated from an open quantum systems formalism \cite{kayman2}.
\end{itemize}
Rather, the goal of this exercise is to provide a proof of principle, showing that hydrodynamics with spin in non-equilibrium combined with a dynamical hadronization process such as coalescence, has the potential to account for the phenomenological puzzles described earlier.
Specifically
\begin{itemize}
\item Vector mesons coalescing in a vortical medium combine two spin $1/2$ and an integer angular momentum into a spin $1$ state, whereas spin $1/2$ baryons coalescing in a vortical medium combine three spin $1/2$ and an integer angular momentum into a spin $1/2$ state.  The latter requires a lot more cancellations (some exact due to spin-statistics theorem), which suppresses polarization.
\item The concrete phenomenological picture implied by spin-vorticity non-equilibrium assumes \cite{kayman2} a global (perpendicular to reaction plane) vorticity formed in the initial state and a local vorticity formed on the time-scale of elliptic flow.   It is reasonable to expect the latter to be less equilibrated with spin than the former.    We note that if angular momentum vorticity is used \cite{florkbook} (vorticity of enthalpy $\times$ flow), it is non-zero also initially due to density gradients, but it is out of phase with the final vorticity (for the same reason as elliptic flow is out of phase with the initial density ellipticity).

Thus the interaction between the different vorticities and polarizations alters the local dependence on the azimuthal angle.
\end{itemize}
Given the large number of parameters used here (and the fact that \cite{FLO} was not tuned to our model in mind), we do not aim to prove that our model is the {\em ultimate} explanation.

Rather, we aim to show how spin hydrodynamics can be phenomenologically probed, and how the parameters above can be quantitatively determined, through a particular observable, namely the azimuthal dependence of both the baryon polarization and the experimentally acessible spin density matrix elements of vector mesons.

We understand that this observable is problematic as it is prone to systematic errors, and that so far only a preliminary measurement exists \cite{qm2018}. However, we hope to demonstrate that the phenomenological potential of such studies motivates their futher development.   A confirmation that off-equilibrium density matrix elements for vector mesons are non-zero and strongly depend on the azimuthal angle would provide evidence supporting our model.
\section{The model}	
\subsection{The coordinate systems \label{expnonrel}}
We begin by describing the coordinate system, shown in Fig. \ref{plot1}. 
 $\theta^*,\phi^*$ refer to the position of the daughter particle within the rest frame of the polarized parent, while $\phi_r$ and $\theta_r$ (where spacetime rapidity is typically used instead) refer to the position of the parent particle with respect to laboratory frame.  Any ensemble average must therefore be obtained in terms of both the daughter decay and the parent distribution.

 The first angle $\phi^*$ is measured with respect to the polarization axis of the parent ($\phi^*=0$ in the polarization direction), while $\phi_r$ is measured with respect to the impact parameter ($\phi_r=0$ corresponds to the maximum of the elliptic flow, with $\pi/2$ relative to $\phi^*=0$).
The second angle, event-by-event, is provided by pseudo-rapidity $\eta$ (Fig. \ref{plot1})\footnote{This notation is opposite to what is usually used in the polarization literature. For consistency with the thermal model used later in this paper $\theta$ and $\phi$ are also exchanged compared to \cite{kayman1}.}
The $\theta^*$ axis is defined relative to the beam, with $\theta^*=0$ corresponding to the pseudo-rapidity of the mother particle, while $\theta_r$ corresponding to the pseudo-rapidity of the daughter particle.

Therefore, for a quantitative estimate, the coalescence equations governing hadron production need to be convoluted with a thermal model distribution (described in subsection \ref{subblast}) to obtain the full momentum rapidity, $p_T$ and angle dependence, considering that
\begin{equation}
\label{convol}
\frac{dN}{d\theta^* d\phi^*} =\int p_T dp_T d y d\phi^* \left(\frac{dN_{parent}}{y dy dp_T d \phi^*}\right)_{parent} \frac{\partial \left( y,p_T,\phi^* \right)}{\partial \left( \theta_r,\phi_r \right)} W(\theta^*-\theta_r,\phi^*-\phi_r)
\end{equation}
where $dN_{parent}/y dy dp_T d \phi^*$ is the distribution in each rapidity, $y,\phi^*$ cells (incorporating longitudinal and elliptic flow) and the Jacobian is the transformation to the rest frame where \eqref{dis} is defined.

In the fully relativistic limit in transverse space, the Jacobian is a very complicated object. However, if the transformation is non-relativistic, it simplifies considerably, since angles are not affected by boosts and the azimuthal dependence of  particle abundances remains constant.  Therefore
\begin{equation}
\frac{\partial \left( \eta,p_T,\phi^* \right)}{\partial \left( \theta_r,\phi_r \right)} \simeq  \underbrace{\frac{d\theta^*}{d\theta_r}}_{\sim const.} \times \frac{d \theta_r}{d\eta}  \eqcomma   \eta \simeq \tanh^{-1}\cos \left( \frac{\theta_r}{2} \right)
\end{equation}
Thus, for an approximately boost invariant system where one observes a small slice in rapidity, it is sufficient to multiply the previous expressions by a weight factor, usually $\sim \order{1}$
\[\   \frac{d\eta}{d\theta_r} = - \frac{\sqrt{1-\eta^{2}}}{2\left(1-\eta^{2} \right)} 
\]
Since the model in \cite{bro,FLO} was not tuned for coalescence and the relevant observables are not yet published, we believe such a more precise quantitative comparison is premature.
\subsection{The blast wave model \label{subblast}}
As a simplified quantitative model for our medium we use the blast wave model \cite{FLO}, which we briefly review here.
The blast wave model applied in heavy-ion collisions assumes the creation of a medium where the particles are formed in local thermal equilibrium, and expand in both longitudinal and transverse directions \cite{Ristea}. The longitudinal expansion is defined as boost-invariant, and the transverse expansion is described using elliptic fluid flow parametrization, allowing for the inclusion of the elliptical flow effects observed in the experimental transverse momentum spectra \cite{FLO}.

\begin{figure}[h]
        \begin{center}
     \epsfig{width=0.75\textwidth,figure=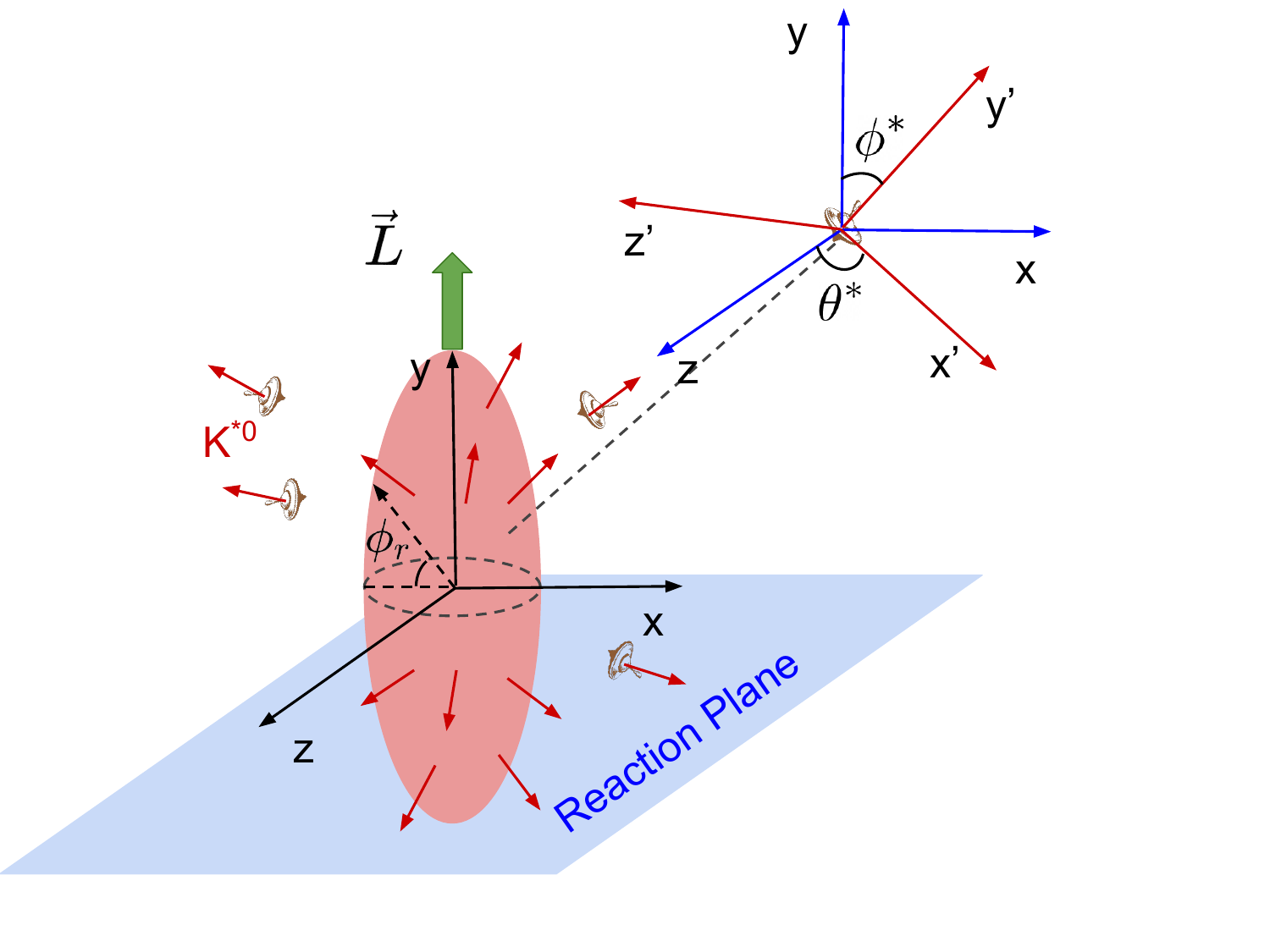}
          \caption{We show a representation of two frames: decay frame, parametrized by the angles $\theta^*$ and $\phi^*$ and the laboratory frame, defined by the angles $\theta_r$ and $\phi_r$   Neglecting relativistic corrections, $\theta_r$ follows the reaction plane (global polarization) and $\phi_r$ follows the longitudinal direction in accordance to Sec.\ref{expnonrel}
                \label{plot1}}
        \end{center}
\end{figure}
Following Ref.\cite{FLO}, employing anizotropic fluid flow parametrization incorporated in the blast wave model we determine the fluid vorticity required for our study. This provides approximate longitudinal component of the vortical structures resulting from elliptical fireball shape in the $x-y$ plane as represented in Fig. \ref{plot1}. The $x$ and $y$ components of the freeze-out hypersurface points are given by (see Sec.\ref{expnonrel} and Fig. \ref{plot1})
\begin{equation}
    \begin{array}{c}
    x = r_{max}\sqrt{1-\epsilon}\cos\phi_r\\
    y = r_{max}\sqrt{1+\epsilon}\sin\phi_r
    \end{array}
    \label{para1}
\end{equation}
where $r_{max}$ and $\epsilon$ are model parameters, and the asymmetric flow velocity is given by:
\begin{equation}
    u^{\mu} = \left(\frac{t}{N},\frac{x}{N}\sqrt{1+\delta},\frac{y}{N}\sqrt{1-\delta},\frac{z}{N}\right)
    \label{flow}
\end{equation}
The normalization parameter in the above equation is:
\begin{equation}
    N = \sqrt{\tau^2-(x^2-y^2)\delta}
\end{equation}
with the proper time expressed as:
\begin{equation}
    \tau^2 = t^2-x^2-y^2-z^2
\end{equation}
We define $z$ as ($\eta$ is related to $\theta_r$, see Sec.\ref{expnonrel})
\begin{equation}
	z = \sqrt{\tau^2_f+x^2+y^2}\sinh\eta 
        \label{para2}
\end{equation}
In Figure \ref{plot1}, all the angles used in this work are clearly depicted. The angle $\phi_r$ is the angle in the $x-y$ plane from the blast wave model, Eq. \eqref{para1} and the angles $\phi^*$ and $\theta^*$ represent the angles between the parent particle's spin and the momentum of the daughter particle.

The $\delta$ and $\epsilon$ parameters describe the non-trivial effects in the $x-y$ plane in non-central collisions, extracted from a fit to 130 GeV spectra and elliptic flow data in \cite{bro,FLO}. Now, we can determine the thermal vorticity in the following form:
\begin{equation} 
    \Omega_{\mu,\nu} = -\frac{1}{2T}(\partial_{\mu} u_\nu - \partial_\nu u_\mu) - \frac{1}{2T^2}(u_\mu\partial_\nu T - u_\nu\partial_\mu T).
    \label{bwvor}
\end{equation}
The spatial vorticity component is defined as:
\begin{equation}
    \Omega_i = \frac{1}{2}\epsilon_{ijk}\Omega_{jk}
\end{equation}
where $i, j, k = 1, 2, 3$. To obtain the explicit value of vorticity, we need to utilize the fireball parametrization equations \eqref{para1} and \eqref{para2} together with the flow velocity \eqref{flow} in the thermal vorticity definition \eqref{bwvor}. Thus, we obtain the following expressions:

\begin{equation}
    \begin{array}{c}
    \Omega_1 = -\frac{yz}{TN^3}\left(\sqrt{1+\delta}-1+\delta\right)\\
    \Omega_2 = \frac{xz}{TN^3}\left(1+\delta-\sqrt{1+\delta}\right)\\
    \Omega_3 = \frac{xy\sqrt{1-\delta^2}}{TN^3}\left(\sqrt{1+\delta}-\sqrt{1-\delta}\right).
    \label{vort}
    \end{array}
\end{equation}
In this way, we can connect the thermal vorticity developed in \cite{FLO} and the coalescence model for vector mesons \cite{kayman2}, and extend this formalism to include baryons. This is important because it allows us to examine the influence of vorticity on the alignment factor and particle polarization within the same framework.
\subsection{Production of Baryons and Vector Mesons}

Using the coalescence model, we can study the effects of vortices in the quark-gluon plasma (QGP) on vector mesons with spin 1, such as $\phi$ and $K^{*}$, and compare these effects to those on baryons with spin 1/2.  We calculate the density matrix coeffients for vector mesons
\begin{equation}
	\left( \hat{\rho}^M \right)^{L}_{m',m'''}= \int_0^{2\pi} P_W(\nu) d\nu \sum_{m'' , m}P^{2}_L(\omega) \tilde{C}^{L}_{m^{''},m} d^{j}_{m' m'' }(\nu)\bra{j,m''}\rho^{1}(\Omega)\rho^{2} (\Omega)\ket{j,m}\left[d^{j}_{m m''}(\nu)\right]^{-1}
	\label{coalmeson}
	\end{equation}
and for baryons  
\begin{equation}
\label{coalbaryon}
    \left( \hat{\rho}^B \right)^{L}_{m',m'''}= \int_0^{2\pi} P_W(\nu) d\nu \sum_{m'' , m}P^{2}_L(\omega) \tilde{C}^{L}_{m^{''},m} d^{j}_{m' m'' }(\nu)\bra{j,m''}\rho^{1}(\Omega)\rho^{2} (\Omega)\rho^{3} (\Omega)\ket{j,m}\left[d^{j}_{m m''}(\nu)\right]^{-1}
\end{equation}
where $\Omega$ represents the vorticity created early in the collision, while $\omega$ is the vorticity formed during the hadronization phase, associated with the angular momentum $L$, as described in detail in \cite{kayman1}:
\begin{itemize}
\item $d^{j}_{...}$ rotate the spin direction of quarks to align with the vorticity $\omega$.
\item $C^{...}_{...}$ are  Clebsh-Gordan coefficients. 
\item $P_L(\omega)$ is the probability of a thermalized fluid with vorticity $\omega$ to transfer momentum quantum number $L$ to the hadron spin structure (though this is something not yet well understood, it is in principle calculable from an open quantum system approach).  Note that only vorticity that can be consistently described this way is the one that carries angular momentum, i.e. thermal vorticity.
\item $P_W(\nu-\nu_0)$ is the coalescence Wigner function in spin space, centered around a specific angle, similar to previous coalescence work \cite{fries} but with just spin degrees of freedom included.  As explained in \cite{kayman1} we assume it to be a $\delta$-function as in \cite{xia} (although it can also be smeared with a Gaussian \cite{kayman1})
\begin{equation}
\label{wignerdef}
    P_W(\nu) =  \prod_i^{2(M)3(B)}\delta(\nu_i-\nu_0)
\end{equation}    
 \item $\rho^{1,2,3}$ is given by:
\begin{equation}
	\hat{\rho}^{1,2,3}_{kl} = \frac{1}{Z}\exp\left[\frac{\boldsymbol{\hat{\Omega}}\cdot\boldsymbol{\hat{\sigma}}}{2}\right].
 \label{rhoquark1}
\end{equation}
\end{itemize}
It is important to note that in the case presented here, we do not consider the rotation $U(\phi_r,\theta_r)$. This is because, assuming longitudinal Bjorken flow (where spacetime and momentum rapidities coincide) and in the non-relativistic limit, the effect of the longitudinal expansion of the QGP is already accounted for in the thermal model \cite{FLO}. To illustrate this, we refer to \cite{xia}, where the local spin alignment is considered using the longitudinal expansion of the QGP. We can reproduce the results of \cite{xia} by setting $\theta_r = \Delta\psi$ and using the numerical values $\phi_r = -\pi/2$ and $\Omega/T=1/5$ for $L = 0$. For more details, consult Appendix C in \cite{kayman1}.

We can obtain the alignment factor as a function of rapidity, $\eta$, using the blast wave model to determine the vorticity, while accounting for the effects of the elliptic flow. In equations \eqref{rhoeq1}, \eqref{rhoeq2}, and \eqref{rhoeq3}, $\rho_{00}$ is given for  $L=0,1,2$ respectively. 
\begin{description}
\item[The coefficient] $\left<\left(\rho^M\right)^{L=0}_{00}\left(\Omega_1,\Omega_2,\Omega_3, T\right)\right>$
\begin{equation}
    \left<\left(\rho^M\right)^{L=0}_{00}\left(\Omega_1,\Omega_2,\Omega_3, T\right)\right> = \frac{1}{N_0\left(\Omega _3^2+2 \left(\Omega _1^2+\Omega
   _2^2\right)\right)}\left(\Omega _3^2 \cos ^2(\nu )+2 \sqrt{2} \Omega _1 \Omega _3 \sin (2 \nu )\right.
   \label{rhoeq1}
\end{equation}
\[
     \left.\times\sinh
   ^2\left(\frac{1}{2} \sqrt{\Omega _3^2+2 \left(\Omega _1^2+\Omega
   _2^2\right)}\right)+\cosh \left(\sqrt{\Omega _3^2+2 \left(\Omega _1^2+\Omega
   _2^2\right)}\right) \left(\Omega _3^2 \sin ^2(\nu )+\right.\right.
\]
\[
    \left.\left.+2 \Omega _2^2\right)+2 \Omega
   _1^2 \left(\cos ^2(\nu ) \cosh \left(\sqrt{\Omega _3^2+2 \left(\Omega _1^2+\Omega
   _2^2\right)}\right)+\sin ^2(\nu )\right)\right)
\]
\[
    N_0 = 1+2 \cosh \left(\sqrt{\Omega _3^2+2
   \left(\Omega _1^2+\Omega
   _2^2\right)}\right)
\]
\item[The coefficient] $\left<\left(\rho^M\right)^{L=1}_{00}\left(\Omega_1,\Omega_2,\Omega_3, T\right)\right>$
\begin{equation}
    \left<\left(\rho^M\right)^{L=1}_{00}\left(\Omega_1,\Omega_2,\Omega_3, T\right)\right> = \frac{1}{N_1\left(4 \Omega _3^2+8
   \left(\Omega _1^2+\Omega _2^2\right)\right)}\left(4 \left(\Omega _2^2+\Omega _3^2\right) \sin ^2(\nu )-8 \Omega _1\times\right.
   \label{rhoeq2}
\end{equation}
\[
    \left.\times\left(\sqrt{\Omega _3^2+2 \left(\Omega _1^2+\Omega _2^2\right)} \sin ^2(\nu ) \sinh
   \left(\sqrt{\Omega _3^2+2 \left(\Omega _1^2+\Omega _2^2\right)}\right)+\sqrt{2}
   \Omega _3 \sin (2 \nu )\times\right.\right.
\]
\[
    \left.\left.\times\sinh ^2\left(\frac{1}{2} \sqrt{\Omega _3^2+2 \left(\Omega
   _1^2+\Omega _2^2\right)}\right)\right)+2 \Omega _3 \sqrt{2 \Omega _3^2+4 \left(\Omega
   _1^2+\Omega _2^2\right)} \sin (2 \nu )\times \right.
\]
\[
    \left.\times\sinh \left(\sqrt{\Omega _3^2+2 \left(\Omega
   _1^2+\Omega _2^2\right)}\right)+\cosh \left(\sqrt{\Omega _3^2+2 \left(\Omega
   _1^2+\Omega _2^2\right)}\right) \left(\Omega _3^2 (\cos (2 \nu )+3)+8 \Omega
   _2^2\right)+\right.
\]
\[
    \left.+4 \Omega _1^2 \left(3 \sin ^2(\nu ) \cosh \left(\sqrt{\Omega _3^2+2
   \left(\Omega _1^2+\Omega _2^2\right)}\right)+2 \cos ^2(\nu )\right)\right)
\]
\[
    N_1 = \frac{1}{\Omega
   _3^2+2 \left(\Omega _1^2+\Omega
   _2^2\right)}\left(-2 \sqrt{\Omega _3^2+2 \left(\Omega
   _1^2+\Omega _2^2\right)} \Omega _1 \sinh
   \left(\sqrt{\Omega _3^2+2
   \left(\Omega _1^2+\Omega
   _2^2\right)}\right)+\right.
\]
\[
    \left.+\left(3 \Omega
   _1^2+5 \Omega _2^2+2 \Omega _3^2\right)
   \cosh \left(\sqrt{\Omega _3^2+2
   \left(\Omega _1^2+\Omega
   _2^2\right)}\right)+3 \Omega
   _1^2+\Omega _2^2+\Omega _3^2\right)
\]
\item[The coefficient] $\left<\left(\rho^M\right)^{L=2}_{00}\left(\Omega_1,\Omega_2,\Omega_3, T\right)\right>$
\begin{equation}
    \left<\left(\rho^M\right)^{L=2}_{00}\left(\Omega_1,\Omega_2,\Omega_3, T\right)\right> = \frac{1}{10 N_2\left(\Omega _3^2+2 \left(\Omega
   _1^2+\Omega _2^2\right)\right)}\left(4 \Omega _1 \left(\left(2 \sqrt{3}-\sqrt{2}\right) \Omega _3 \sin (2 \nu )\times\right.\right. 
   \label{rhoeq3}
\end{equation}
\[
    \left.\left.\times\sinh
   ^2\left(\frac{1}{2} \sqrt{\Omega _3^2+2 \left(\Omega _1^2+\Omega
   _2^2\right)}\right)-2 \sqrt{3 \Omega _3^2+6 \left(\Omega _1^2+\Omega _2^2\right)}
   \cos ^2(\nu ) \sinh \left(\sqrt{\Omega _3^2+2 \left(\Omega _1^2+\Omega
   _2^2\right)}\right)\right)+\right.
\]
\[
    \left.+2 \Omega _1^2 \left(10 \cos ^2(\nu ) \cosh
   \left(\sqrt{\Omega _3^2+2 \left(\Omega _1^2+\Omega _2^2\right)}\right)+\left(7-2
   \sqrt{6}\right) \sin ^2(\nu )\right)+2 \Omega _2^2\times\right. 
\]
\[
    \left.\times\left(\cosh \left(\sqrt{\Omega
   _3^2+2 \left(\Omega _1^2+\Omega _2^2\right)}\right) \left(\left(7-2 \sqrt{6}\right)
   \sin ^2(\nu )+4 \cos ^2(\nu )\right)+6 \cos ^2(\nu )\right)+\right.
\]
\[
    \left.+\Omega _3
   \left(\left(\sqrt{6}-6\right) \sqrt{\Omega _3^2+2 \left(\Omega _1^2+\Omega
   _2^2\right)} \sin (2 \nu ) \sinh \left(\sqrt{\Omega _3^2+2 \left(\Omega _1^2+\Omega
   _2^2\right)}\right)+\right.\right.
\]
\[
    \left.\left.+\Omega _3 \left(\cosh \left(\sqrt{\Omega _3^2+2 \left(\Omega
   _1^2+\Omega _2^2\right)}\right) \left(\left(7-2 \sqrt{6}\right) \sin ^2(\nu )+6 \cos
   ^2(\nu )\right)+4 \cos ^2(\nu )\right)\right)\right)
\]
\[
    N_2 = \frac{1}{5 \left(\Omega _3^2+2 \left(\Omega
   _1^2+\Omega _2^2\right)\right)}\left(-6 \left(\sqrt{2}+\sqrt{3}\right)
   \sqrt{\Omega _3^2+2 \left(\Omega
   _1^2+\Omega _2^2\right)} \Omega _1 \sinh
   \left(\sqrt{\Omega _3^2+2
   \left(\Omega _1^2+\Omega
   _2^2\right)}\right)+\right.
\]
\[
    \left. +\Omega _1^2
   \left(\left(23+2 \sqrt{6}\right) \cosh
   \left(\sqrt{\Omega _3^2+2
   \left(\Omega _1^2+\Omega
   _2^2\right)}\right)-2
   \sqrt{6}+7\right)+\left(\left(17-2
   \sqrt{6}\right) \Omega _2^2+10 \Omega
   _3^2\right)\times \right. 
\]
\[
    \left. \times \cosh
   \left(\sqrt{\Omega _3^2+2
   \left(\Omega _1^2+\Omega
   _2^2\right)}\right)+\left(13+2
   \sqrt{6}\right) \Omega _2^2+5 \Omega
   _3^2\right) 
\]
\end{description}
Note that some combinations of $P_L$ and polarization are forbidden by symmetry (the Pauli-Lubanski decomposition ensures $L=0$ $\chi_{1/2}=0$). For example $L=0$ cannot result in baryon spin alignment.  Nevertheless, we have calculated these cases as a test of our numerical model.

The off-diagonal matrix elements will be
\begin{description}
\item[The coefficient]$\left<r^{L=0}_{1,0}\left(\Omega_1,\Omega_2,\Omega_3, T,\phi_r\right)\right>$
\begin{equation}
    \left<r^{L=0}_{1,0}\left(\Omega_1,\Omega_2,\Omega_3, T,\phi_r\right)\right>=\frac{1}{N_0\left(2 \Omega _1^2+2 \Omega _2^2+\Omega _3^2\right)}\left(\sinh ^2\left(\frac{1}{2} \sqrt{2 \Omega _1^2+2 \Omega _2^2+\Omega _3^2}\right)\times\right.
\end{equation}
\[
    \left.\times\left(2 \sqrt{2} \Omega _1^2 \sin (2 \nu )-\sqrt{2} \Omega _3^2 \sin (2 \nu )-4
   \Omega _3 \Omega _1 \cos (2 \nu )\right)\right)
\]
\item[The coefficient]$\left<r^{L=1}_{1,0}\left(\Omega_1,\Omega_2,\Omega_3, T,\phi_r\right)\right>$
\begin{equation}
    \left<r^{L=1}_{1,0}\left(\Omega_1,\Omega_2,\Omega_3, T,\phi_r\right)\right>=\frac{1}{\left(2 \Omega _3^2+4 \left(\Omega _1^2+\Omega _2^2\right)\right)}\left(2 \sqrt{2} \Omega _1^2 \sin (2 \nu )-\sqrt{2} \Omega _2^2 \sin (2 \nu )+\right.
\end{equation}
\[
    \left.-\sqrt{2}
   \Omega _3^2 \sin (2 \nu )+8 \Omega _1 \Omega _3 \cos (2 \nu ) \sinh
   ^2\left(\frac{1}{2} \sqrt{\Omega _3^2+2 \left(\Omega _1^2+\Omega
   _2^2\right)}\right)-2 \Omega _3 \sqrt{\Omega _3^2+2 \left(\Omega _1^2+\Omega
   _2^2\right)}\times\right. 
\]
\[
    \left.\times\cos (2 \nu ) \sinh \left(\sqrt{\Omega _3^2+2 \left(\Omega _1^2+\Omega
   _2^2\right)}\right)+4 \Omega _1 \sqrt{2 \Omega _3^2+4 \left(\Omega _1^2+\Omega
   _2^2\right)} \sin (\nu ) \cos (\nu )\times\right. 
\]
\[
    \left.\times\sinh \left(\sqrt{\Omega _3^2+2 \left(\Omega
   _1^2+\Omega _2^2\right)}\right)+\sqrt{2} \left(\Omega _3^2-6 \Omega _1^2\right) \sin
   (\nu ) \cos (\nu ) \cosh \left(\sqrt{\Omega _3^2+2 \left(\Omega _1^2+\Omega
   _2^2\right)}\right)\right)
\]
\item[The coefficient]$\left<r^{L=2}_{1,0}\left(\Omega_1,\Omega_2,\Omega_3, T,\phi_r\right)\right>$
\begin{equation}
    \left<r^{L=2}_{1,0}\left(\Omega_1,\Omega_2,\Omega_3, T,\phi_r\right)\right>=\frac{1}{20 N_2\left(\Omega _3^2+2
   \left(\Omega _1^2+\Omega _2^2\right)\right)}\left(-8 \sqrt{6} \Omega _1 \sqrt{\Omega _3^2+2 \left(\Omega _1^2+\Omega _2^2\right)}
   \sin (2 \nu )\times\right. 
\end{equation}
\[
    \left.\times\sinh \left(\sqrt{\Omega _3^2+2 \left(\Omega _1^2+\Omega
   _2^2\right)}\right)-16 \left(\sqrt{6}-1\right) \Omega _1 \Omega _3 \cos (2 \nu )
   \sinh ^2\left(\frac{1}{2} \sqrt{\Omega _3^2+2 \left(\Omega _1^2+\Omega
   _2^2\right)}\right)+\right.
\]
\[
    \left.+4 \sqrt{21-6 \sqrt{6}} \Omega _3 \sqrt{\Omega _3^2+2 \left(\Omega
   _1^2+\Omega _2^2\right)} \cos (2 \nu ) \sinh \left(\sqrt{\Omega _3^2+2 \left(\Omega
   _1^2+\Omega _2^2\right)}\right)+\sin (2 \nu )\times\right. 
\]
\[
    \left.\times\left(\left(8 \sqrt{3}-14
   \sqrt{2}\right) \Omega _1^2+4 \sqrt{2} \left(3 \Omega _2^2+\Omega
   _3^2\right)+\left(20 \sqrt{2} \Omega _1^2+\left(8 \sqrt{3}-6 \sqrt{2}\right) \Omega
   _2^2+\sqrt{50-8 \sqrt{6}} \Omega _3^2\right)\times\right.\right. 
\]
\[
    \left.\left.\times \cosh \left(\sqrt{\Omega _3^2+2
   \left(\Omega _1^2+\Omega _2^2\right)}\right)\right)\right)
\]
\item[The coefficient]$\left<\alpha^{L=0}_{1,0}\left(\Omega_1,\Omega_2,\Omega_3, T,\phi_r\right)\right>$
\begin{equation}
    \left<\alpha^{L=0}_{1,0}\left(\Omega_1,\Omega_2,\Omega_3, T,\phi_r\right)\right>= -\frac{2 \Omega _2 \sinh \left(\sqrt{\Omega _3^2+2 \left(\Omega _1^2+\Omega
   _2^2\right)}\right)}{N_0\sqrt{\Omega _3^2+2 \left(\Omega _1^2+\Omega _2^2\right)}}
\end{equation}
\item[The coefficient]$\left<\alpha^{L=1}_{1,0}\left(\Omega_1,\Omega_2,\Omega_3, T,\phi_r\right)\right>$
\begin{equation}
    \left<\alpha^{L=1}_{1,0}\left(\Omega_1,\Omega_2,\Omega_3, T,\phi_r\right)\right>=\frac{2 \Omega _2}{N_1} \left(\frac{\Omega _1 \left(\cosh \left(\sqrt{\Omega _3^2+2 \left(\Omega
   _1^2+\Omega _2^2\right)}\right)-1\right)}{\Omega _3^2+2 \left(\Omega _1^2+\Omega
   _2^2\right)}\right.
\end{equation}
\[
    \left.-\frac{\sinh \left(\sqrt{\Omega _3^2+2 \left(\Omega _1^2+\Omega
   _2^2\right)}\right)}{\sqrt{\Omega _3^2+2 \left(\Omega _1^2+\Omega
   _2^2\right)}}\right)
\]
\item[The coefficient]$\left<\alpha^{L=2}_{1,0}\left(\Omega_1,\Omega_2,\Omega_3, T,\phi_r\right)\right>$
\begin{equation}
    \left<\alpha^{L=2}_{1,0}\left(\Omega_1,\Omega_2,\Omega_3, T,\phi_r\right)\right>= \frac{4}{5N_2} \Omega _2 \sinh \left(\frac{1}{2} \sqrt{\Omega _3^2+2 \left(\Omega
   _1^2+\Omega _2^2\right)}\right)\times 
\end{equation}
\[
    \times\left(\frac{\sqrt{21-6 \sqrt{6}} \Omega _1 \sinh
   \left(\frac{1}{2} \sqrt{\Omega _3^2+2 \left(\Omega _1^2+\Omega
   _2^2\right)}\right)}{\Omega _3^2+2 \left(\Omega _1^2+\Omega
   _2^2\right)}-\frac{\left(\sqrt{6}-1\right) \cosh \left(\frac{1}{2} \sqrt{\Omega
   _3^2+2 \left(\Omega _1^2+\Omega _2^2\right)}\right)}{\sqrt{\Omega _3^2+2 \left(\Omega
   _1^2+\Omega _2^2\right)}}\right)
\]
\item[The coefficient] $\left<r^{L=0}_{1,-1}\left(\Omega_1,\Omega_2,\Omega_3, T,\phi_r\right)\right>$
\begin{equation}
    \left<r^{L=0}_{1,-1}\left(\Omega_1,\Omega_2,\Omega_3, T,\phi_r\right)\right> =\frac{1}{N_0\left(\Omega _3^2+2 \left(\Omega
   _1^2+\Omega _2^2\right)\right)} \left(\sinh ^2\left(\frac{1}{2} \sqrt{\Omega _3^2+2 \left(\Omega _1^2+\Omega
   _2^2\right)}\right)\times\right. 
   \label{r1m1eq1}
\end{equation}
\[
    \left.\times \left(\Omega _3^2 \sin ^2(\nu )+\sqrt{2} \Omega _3 \Omega _1 \sin
   (2 \nu )+2 \Omega _1^2 \cos ^2(\nu )-2 \Omega _2^2\right)\right)
\]
\item[The coefficient] $\left<r^{L=1}_{1,-1}\left(\Omega_1,\Omega_2,\Omega_3, T,\phi_r\right)\right>$
\begin{equation}
   \left<r^{L=1}_{1,-1}\left(\Omega_1,\Omega_2,\Omega_3, T,\phi_r\right)\right> = \frac{1}{4 N_1\left(\Omega _3^2+2
   \left(\Omega _1^2+\Omega _2^2\right)\right){}^{3/2}}\left(8 \Omega _1^3 \cos ^2(\nu )\times\right. 
\end{equation}
\[
    \left.\times \sinh \left(\sqrt{\Omega _3^2+2 \left(\Omega
   _1^2+\Omega _2^2\right)}\right)+\Omega _3^2 \left(\sqrt{2} \Omega _3 \sin (2 \nu )
   \sinh \left(\sqrt{\Omega _3^2+2 \left(\Omega _1^2+\Omega
   _2^2\right)}\right)\right.\right.
\]
\[
    \left.\left.-\sqrt{\Omega _3^2+2 \left(\Omega _1^2+\Omega _2^2\right)}
   \left(\sin ^2(\nu ) \cosh \left(\sqrt{\Omega _3^2+2 \left(\Omega _1^2+\Omega
   _2^2\right)}\right)+2 \cos ^2(\nu )\right)\right)+\right.
\]
\[
    \left.+2 \Omega _1^2 \left(\sqrt{2} \Omega
   _3 \sin (2 \nu ) \sinh \left(\sqrt{\Omega _3^2+2 \left(\Omega _1^2+\Omega
   _2^2\right)}\right)+\sqrt{\Omega _3^2+2 \left(\Omega _1^2+\Omega _2^2\right)}
   \left(\cos (2 \nu )\right.\right.\right.
\]
\[
    \left.\left.\left.-3 \cos ^2(\nu ) \cosh \left(\sqrt{\Omega _3^2+2 \left(\Omega
   _1^2+\Omega _2^2\right)}\right)\right)\right)+2 \Omega _1 \left(\Omega _3 \sqrt{2
   \Omega _3^2+4 \left(\Omega _1^2+\Omega _2^2\right)} \sin (2 \nu )\right.\right.
\]
\[
    \left.\left.+2 \left(2 \Omega
   _2^2+\Omega _3^2\right) \cos ^2(\nu ) \sinh \left(\sqrt{\Omega _3^2+2 \left(\Omega
   _1^2+\Omega _2^2\right)}\right)-2 \Omega _3 \sqrt{2 \Omega _3^2+4 \left(\Omega
   _1^2+\Omega _2^2\right)} \sin (\nu ) \cos (\nu )\times\right.\right. 
\]
\[
    \left.\left.\times \cosh \left(\sqrt{\Omega _3^2+2
   \left(\Omega _1^2+\Omega _2^2\right)}\right)\right)-\Omega _2^2 \left(\sqrt{\Omega
   _3^2+2 \left(\Omega _1^2+\Omega _2^2\right)}\times\right.\right. 
\]
\[
    \left.\left.\times\left(\cos (2 \nu )+2 \cosh
   \left(\sqrt{\Omega _3^2+2 \left(\Omega _1^2+\Omega _2^2\right)}\right)+1\right)\right.\right.
\]
\[
    \left.\left.-4
   \sqrt{2} \Omega _3 \sin (\nu ) \cos (\nu ) \sinh \left(\sqrt{\Omega _3^2+2
   \left(\Omega _1^2+\Omega _2^2\right)}\right)\right)\right)
\]
\item[The coefficient] $\left<r^{L=2}_{1,-1}\left(\Omega_1,\Omega_2,\Omega_3, T,\phi_r\right)\right>$
\begin{equation}
    \left<r^{L=2}_{1,-1}\left(\Omega_1,\Omega_2,\Omega_3, T,\phi_r\right)\right>=\frac{1}{20 N_2\left(\Omega _3^2+2 \left(\Omega _1^2+\Omega
   _2^2\right)\right){}^{3/2}}\left(-8 \Omega _1^3 \left(\sqrt{3} \cos (2 \nu )+3 \sqrt{2}\right)\times\right.
\end{equation}
\[
    \left.\times  \sinh
   \left(\sqrt{\Omega _3^2+2 \left(\Omega _1^2+\Omega _2^2\right)}\right)-4 \Omega _1
   \left(2 \Omega _2^2 \left(\sqrt{3} \cos (2 \nu )+3 \sqrt{2}\right) \sinh
   \left(\sqrt{\Omega _3^2+2 \left(\Omega _1^2+\Omega _2^2\right)}\right)+\right.\right.
\]
\[
    \left.\left.+\Omega _3
   \left(\left(\sqrt{2}-2 \sqrt{3}\right) \sqrt{\Omega _3^2+2 \left(\Omega _1^2+\Omega
   _2^2\right)} \sin (2 \nu ) \sinh ^2\left(\frac{1}{2} \sqrt{\Omega _3^2+2 \left(\Omega
   _1^2+\Omega _2^2\right)}\right)+\right.\right.\right.
\]
\[
    \left.\left.\left. +\Omega _3 \left(\sqrt{3} \cos (2 \nu )+3
   \sqrt{2}\right) \sinh \left(\sqrt{\Omega _3^2+2 \left(\Omega _1^2+\Omega
   _2^2\right)}\right)\right)\right)+\Omega _1^2 \left(2 \left(\sqrt{6}-6\right) \Omega
   _3 \sin (2 \nu )\times\right.\right. 
\]
\[
    \left.\left.\times \sinh \left(\sqrt{\Omega _3^2+2 \left(\Omega _1^2+\Omega
   _2^2\right)}\right)+2 \sqrt{\Omega _3^2+2 \left(\Omega _1^2+\Omega _2^2\right)}
   \left(\left(5 \cos (2 \nu )+2 \sqrt{6}+8\right)\times\right.\right.\right. 
\]
\[
    \left.\left.\left.\times\cosh \left(\sqrt{\Omega _3^2+2
   \left(\Omega _1^2+\Omega _2^2\right)}\right)+\left(2 \sqrt{6}-7\right) \cos ^2(\nu
   )\right)\right)+\Omega _2^2 \left(2 \left(\sqrt{6}-6\right) \Omega _3 \sin (2 \nu )\times\right.\right.
\]
\[
    \left.\left.\times\sinh \left(\sqrt{\Omega _3^2+2 \left(\Omega _1^2+\Omega
   _2^2\right)}\right)+\sqrt{\Omega _3^2+2 \left(\Omega _1^2+\Omega _2^2\right)}
   \left(\left(\left(2 \sqrt{6}-3\right) \cos (2 \nu )+2 \sqrt{6}+1\right)\times\right.\right.\right. 
\]
\[
    \left.\left.\left.\times\cosh
   \left(\sqrt{\Omega _3^2+2 \left(\Omega _1^2+\Omega _2^2\right)}\right)+6 \cos (2 \nu
   )+4 \sqrt{6}+8\right)\right)+\frac{1}{2} \Omega _3^2 \left(2 \left(\sqrt{6}-6\right)
   \Omega _3 \sin (2 \nu )\times\right.\right.
\]
\[
    \left.\left. \times\sinh \left(\sqrt{\Omega _3^2+2 \left(\Omega _1^2+\Omega
   _2^2\right)}\right)+\sqrt{\Omega _3^2+2 \left(\Omega _1^2+\Omega _2^2\right)}
   \left(\left(\left(2 \sqrt{6}-1\right) \cos (2 \nu )+6 \sqrt{6}+1\right)\times\right.\right.\right.
\]
\[
    \left.\left. \left.\times\cosh
   \left(\sqrt{\Omega _3^2+2 \left(\Omega _1^2+\Omega _2^2\right)}\right)+4 (\cos (2 \nu
   )+2)\right)\right)\right)
\]
\item[The coefficient] $\left<\alpha^{L=0}_{1,-1}\left(\Omega_1,\Omega_2,\Omega_3, T,\phi_r\right)\right>$
\begin{equation}
    \left<\alpha^{L=0}_{1,-1}\left(\Omega_1,\Omega_2,\Omega_3, T,\phi_r\right)\right>=-\frac{2 \Omega _2 \sinh ^2\left(\frac{1}{2} \sqrt{\Omega _3^2+2 \left(\Omega
   _1^2+\Omega _2^2\right)}\right) \left(\sqrt{2} \Omega _3 \sin (\nu )+2 \Omega _1 \cos
   (\nu )\right)}{N_0\left(\Omega _3^2+2 \left(\Omega _1^2+\Omega _2^2\right)\right)}
\end{equation}

\item[The coefficient] $\left<\alpha^{L=1}_{1,-1}\left(\Omega_1,\Omega_2,\Omega_3, T,\phi_r\right)\right>$
\begin{equation}
    \left<\alpha^{L=1}_{1,-1}\left(\Omega_1,\Omega_2,\Omega_3, T,\phi_r\right)\right>=\frac{\Omega _2 \cos (\nu )}{N_1}\times 
\end{equation}
\[
    \times\left(\frac{\Omega _1 \left(\cosh \left(\sqrt{\Omega _3^2+2
   \left(\Omega _1^2+\Omega _2^2\right)}\right)-1\right)}{\Omega _3^2+2 \left(\Omega
   _1^2+\Omega _2^2\right)}-\frac{\sinh \left(\sqrt{\Omega _3^2+2 \left(\Omega
   _1^2+\Omega _2^2\right)}\right)}{\sqrt{\Omega _3^2+2 \left(\Omega _1^2+\Omega
   _2^2\right)}}\right)
\]
\item[The coefficient] $\left<\alpha^{L=2}_{1,-1}\left(\Omega_1,\Omega_2,\Omega_3, T,\phi_r\right)\right>$
\begin{equation}
    \left<\alpha^{L=2}_{1,-1}\left(\Omega_1,\Omega_2,\Omega_3, T,\phi_r\right)\right>=\frac{1}{N_2\left(5 \left(\Omega _3^2+2 \left(\Omega
   _1^2+\Omega _2^2\right)\right)\right)}\left(\Omega _2 \left(2 \sinh ^2\left(\frac{1}{2} \sqrt{\Omega _3^2+2 \left(\Omega
   _1^2+\Omega _2^2\right)}\right)\times\right.\right. 
\end{equation}
\[
    \left.\left.\times\left(2 \sqrt{5-2 \sqrt{6}} \Omega _3 \sin (\nu )+5
   \Omega _1 \cos (\nu )\right)-\sqrt{21-6 \sqrt{6}} \sqrt{\Omega _3^2+2 \left(\Omega
   _1^2+\Omega _2^2\right)}\times\right.\right. 
\]
\[
    \left.\left.\times\cos (\nu ) \sinh \left(\sqrt{\Omega _3^2+2 \left(\Omega
   _1^2+\Omega _2^2\right)}\right)\right)\right)
\]
\end{description}
Finally, from equation \eqref{coalbaryon}, we can determine the polarization of baryons for differents values of angular momentum $L = 0$ and $L =1$. The longitudinal polarization $\chi^{L}_{1/2}$ is defined as $(\left(\rho^B\right)^{L}_{\frac{1}{2},\frac{1}{2}}-\left(\rho^B\right)^{L}_{-\frac{1}{2},-\frac{1}{2}})/(\left(\rho^B\right)^{L}_{\frac{1}{2},\frac{1}{2}}+\left(\rho^B\right)^{L}_{-\frac{1}{2},-\frac{1}{2}})$ and is given in equations \eqref{rhoba1} and \eqref{rhoba2}.
\begin{description}
\item[The coefficient] $ \left<\chi^{L=0}_{1/2}\left(\Omega_1,\Omega_2,\Omega_3, T,\nu\right)\right>$
\begin{equation}
    \left<\chi^{L=0}_{1/2}\left(\Omega_1,\Omega_2,\Omega_3, T,\nu\right)\right> = \frac{\tanh \left(\frac{\left|\boldsymbol{\Omega}\right|}{2} \right)
   \left(\Omega _3 \cos (\nu )-\Omega _1 \sin (\nu )\right)}{\left|\boldsymbol{\Omega}\right|}
   \label{rhoba1}
\end{equation}
\[
    \left|\boldsymbol{\Omega}\right| = \sqrt{\Omega _1^2+\Omega _2^2+\Omega _3^2}
\]
\item[The coefficient] $ \left<\chi^{L=1}_{1/2}\left(\Omega_1,\Omega_2,\Omega_3, T,\nu\right)\right>$
\begin{equation}
    \left<\chi^{L=1}_{1/2}\left(\Omega_1,\Omega_2,\Omega_3, T,\nu\right)\right> = \frac{1}{D}\left(\left(\Omega _3 \cos (\nu )-3 \Omega _1 \sin (\nu )\right) \sinh \left(\frac{\left|\boldsymbol{\Omega}\right|}{2} \right)+\right.
   \label{rhoba2}
\end{equation}
\[
    \left.+2 \sqrt{2} \left|\boldsymbol{\Omega}\right| \sin (\nu ) \cosh \left(\frac{\left|\boldsymbol{\Omega}\right|}{2}\right)\right)
\]
\[
    D = 2 \sqrt{2} \Omega _1 \sinh \left(\frac{\left|\boldsymbol{\Omega}\right|}{2} \right)-3 \left|\boldsymbol{\Omega}\right| \cosh
   \left(\frac{\left|\boldsymbol{\Omega}\right|}{2} \right)
\]
\end{description}
The transverse polarization is defined as $\chi^{L=1}_{1/2}\left(\Omega_1,\Omega_2,\Omega_3, T,\nu\right)=\text{Tr}\left[\hat{\sigma_y} \cdot\hat{\rho}^{B}\right]$, and thus:
\begin{description}
\item[The coefficient] $ \left<\chi^{L=0}_{1/2}\left(\Omega_1,\Omega_2,\Omega_3, T,\nu\right)\right>$
\begin{equation}
    \left<\chi^{L=0}_{1/2}\left(\Omega_1,\Omega_2,\Omega_3, T,\nu\right)\right> =-\frac{\Omega _2 \tanh \left(\frac{1}{2} \left|\boldsymbol{\Omega}\right|\right)}{\left|\boldsymbol{\Omega}\right|} 
\end{equation}
\item[The coefficient] $ \left<\chi^{L=1}_{1/2}\left(\Omega_1,\Omega_2,\Omega_3, T,\nu\right)\right>$
\begin{equation}
    \left<\chi^{L=1}_{1/2}\left(\Omega_1,\Omega_2,\Omega_3, T,\nu\right)\right> = \frac{\Omega _2}{2 \sqrt{2} \Omega _1-3 \left|\boldsymbol{\Omega}\right| \coth
   \left(\frac{1}{2} \left|\boldsymbol{\Omega}\right|\right)}
\end{equation}

\end{description}
\section{Results and discussion}
\begin{figure}[h]
		\begin{center}
    \epsfig{width=0.49\textwidth,figure=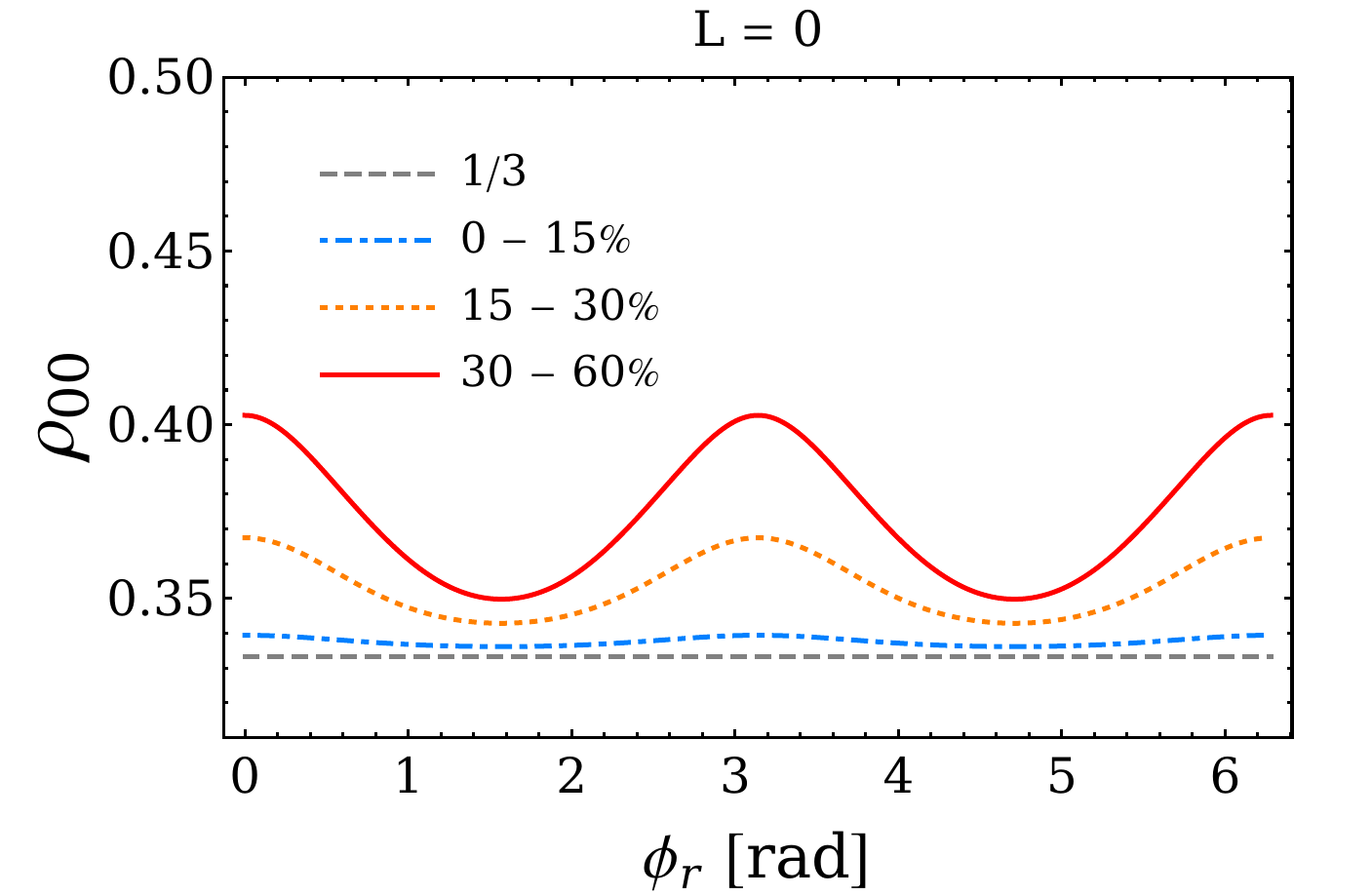}
   \epsfig{width=0.49\textwidth,figure=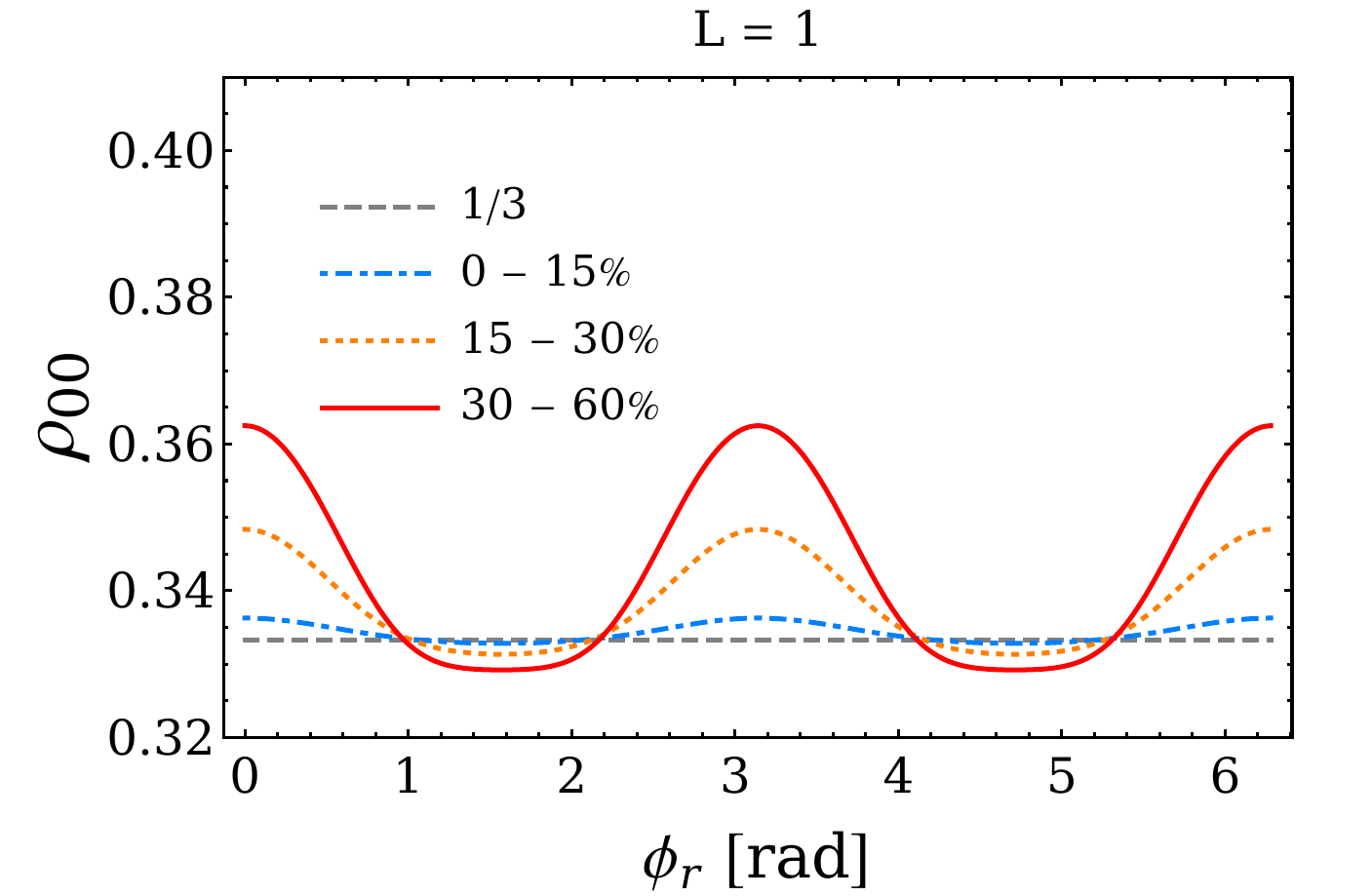}
   \epsfig{width=0.49\textwidth,figure=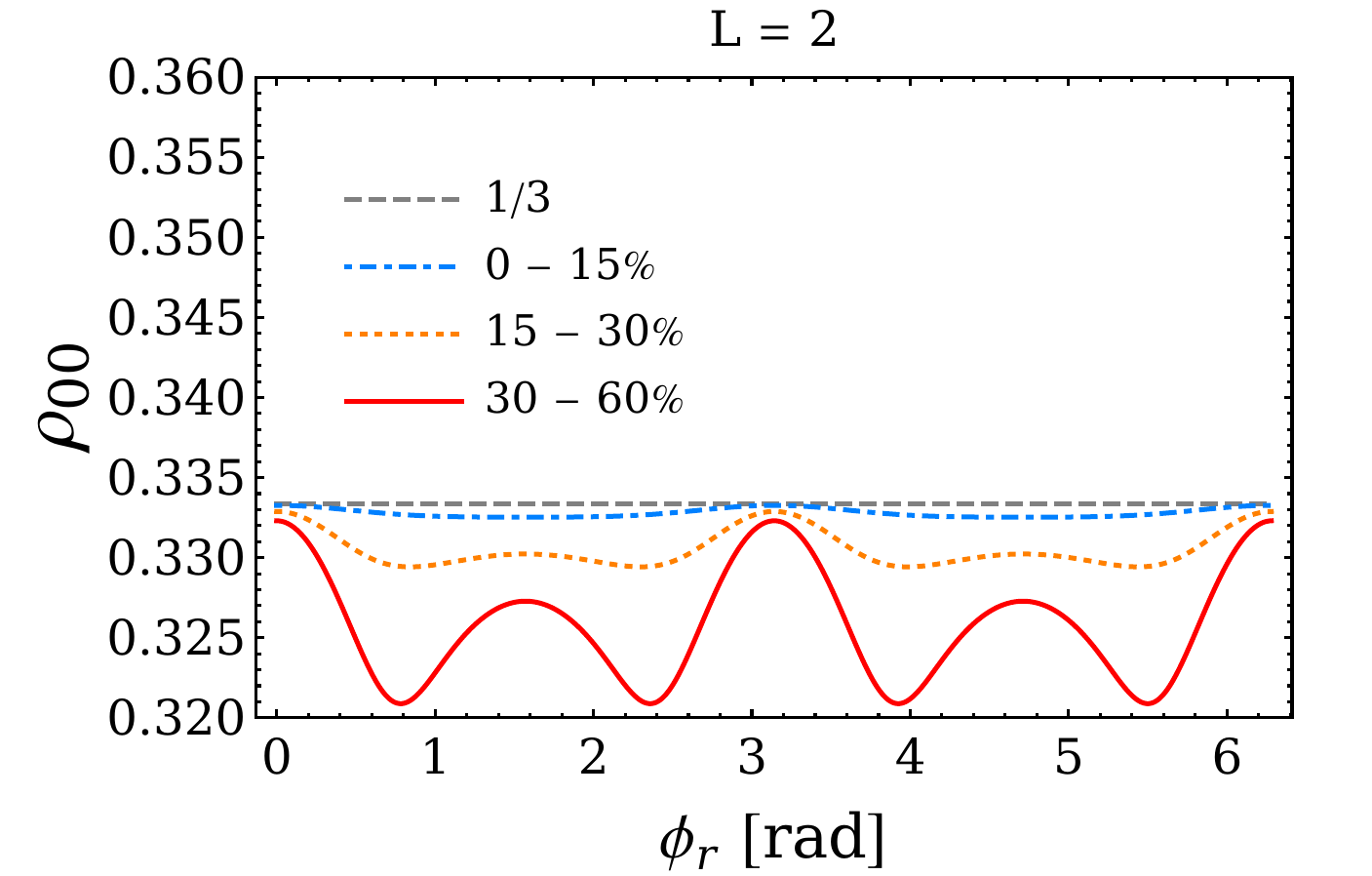}
            \caption{The alignment factor obtained from the coalescence model combined with the blast wave model at 
$\sqrt{s_{NN}} = 130\;\text{GeV}$, relative to azimuthal angle $\phi_r$.\label{rhophi}}
        \end{center}
\end{figure}

\begin{figure}[h]
		\begin{center}
    \epsfig{width=0.49\textwidth,figure=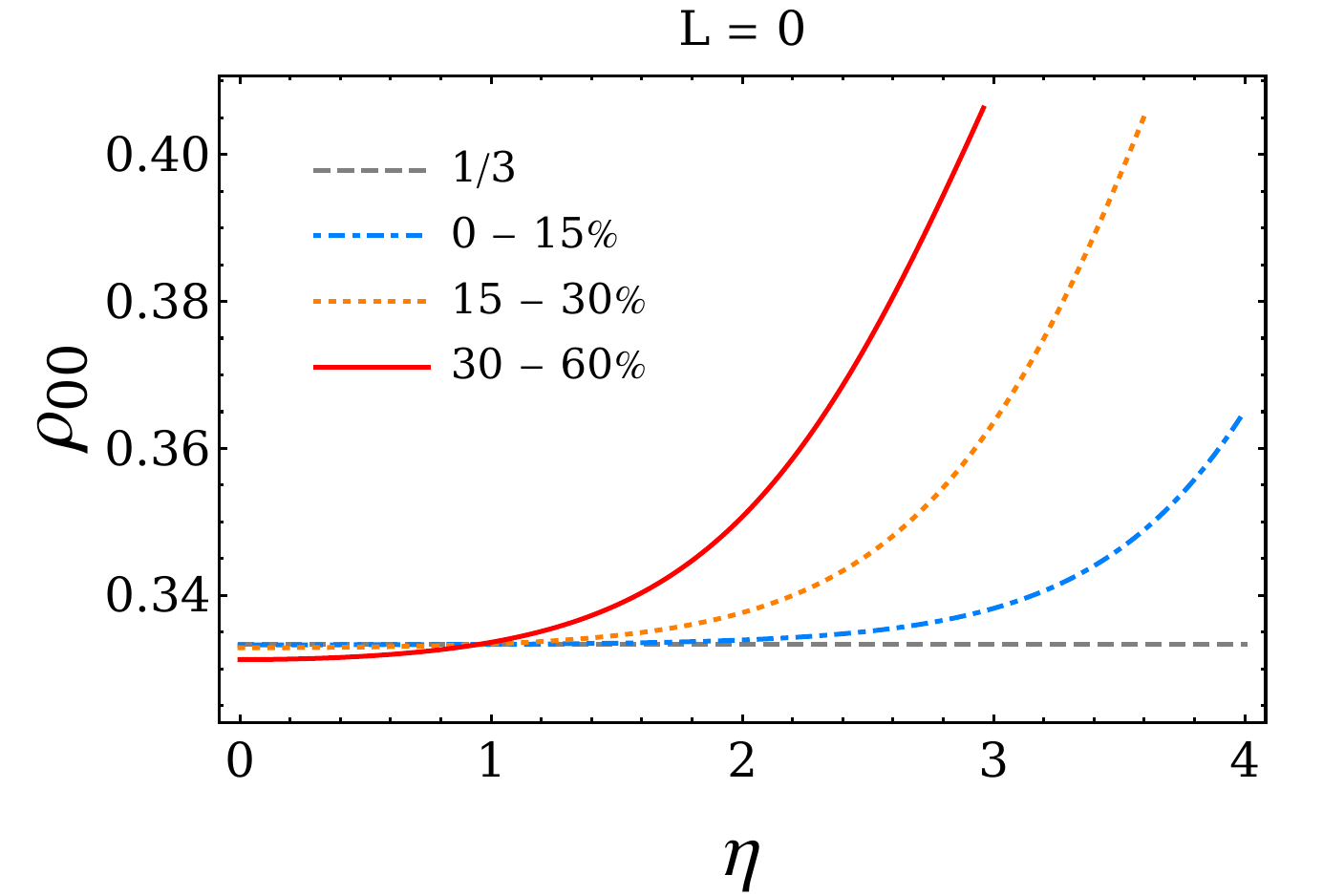}
   \epsfig{width=0.49\textwidth,figure=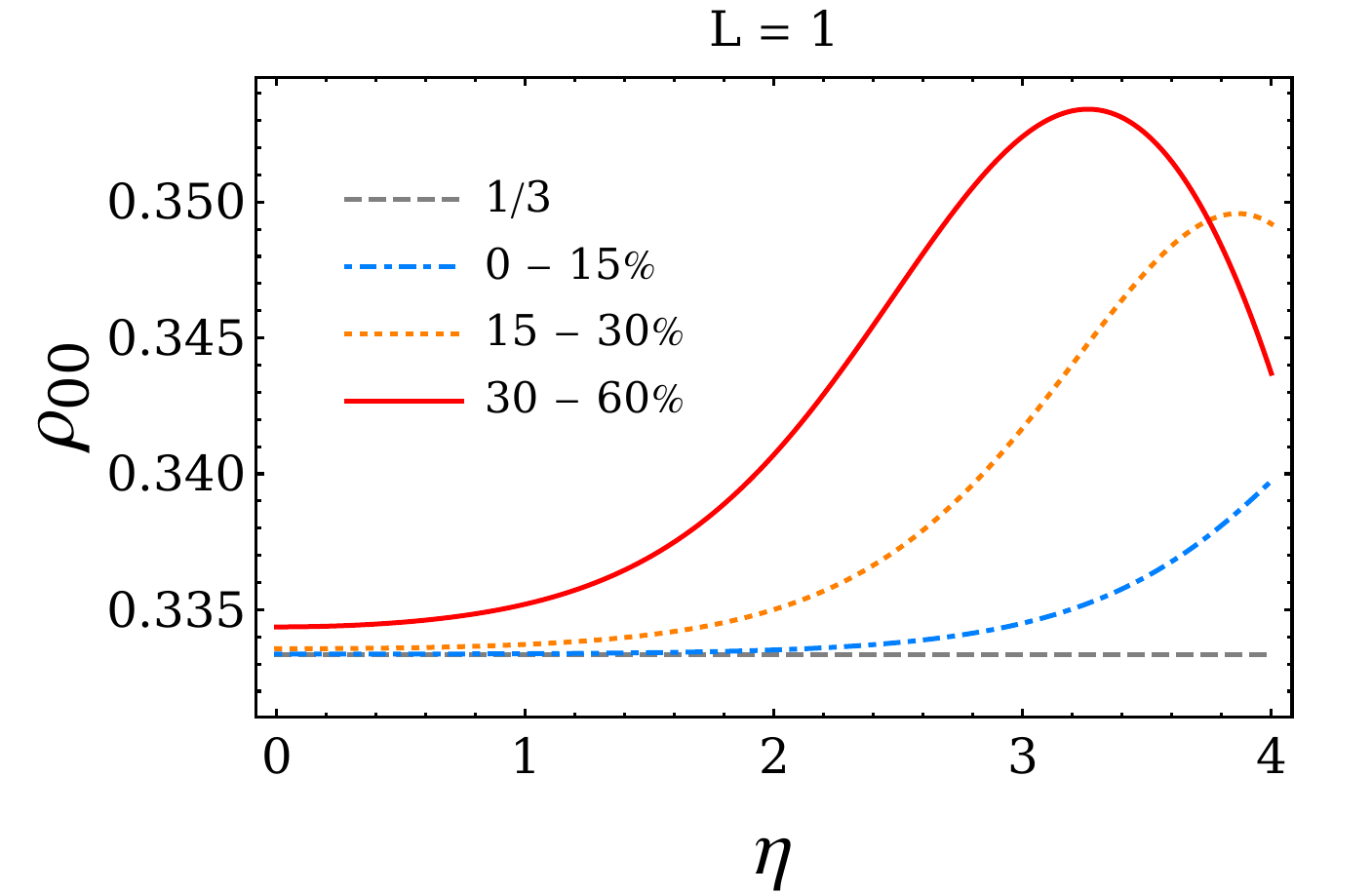}
   \epsfig{width=0.49\textwidth,figure=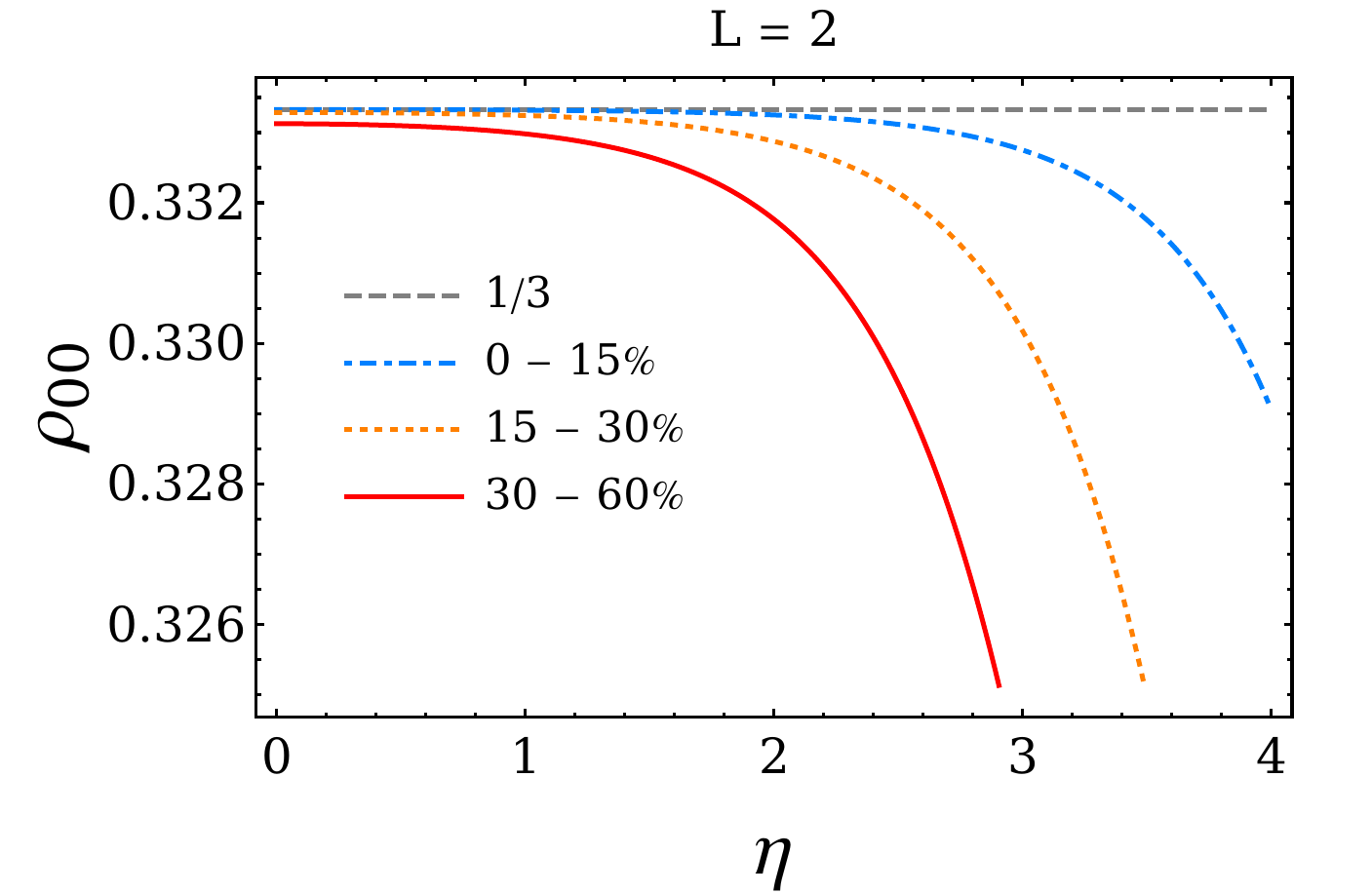}
            \caption{The alignment factor obtained from the coalescence model together with the blast wave model at $\sqrt{s_{NN}} = 130\;\text{GeV}$, as a function of space-time rapidity for different values of angular momentum transferred from the vorticity to the hadron.  Note that as the convolution in \eqref{convol} approaches the Bjorken limit the rapdity distribution of $\rho_{00}$ should follow $\eta$. \label{rhora1}}
        \end{center}
\end{figure} 
From equations \eqref{rhoeq1}, \eqref{rhoeq2}, \eqref{rhoeq3} and vorticity equations defined in \eqref{vort}, we can derive the alignment factor in relation to the azimuthal angle $\phi_r$ for the space-time rapidity range $\left|\eta\right|<4$. In Fig. \ref{rhophi} all angular momentum contributions $(L=0,1,2)$
of alignment factor exhibit an oscillatory behavior, with the amplitude increasing as centrality increases.

\begin{figure}[h]
		\begin{center}
    \epsfig{width=0.49\textwidth,figure=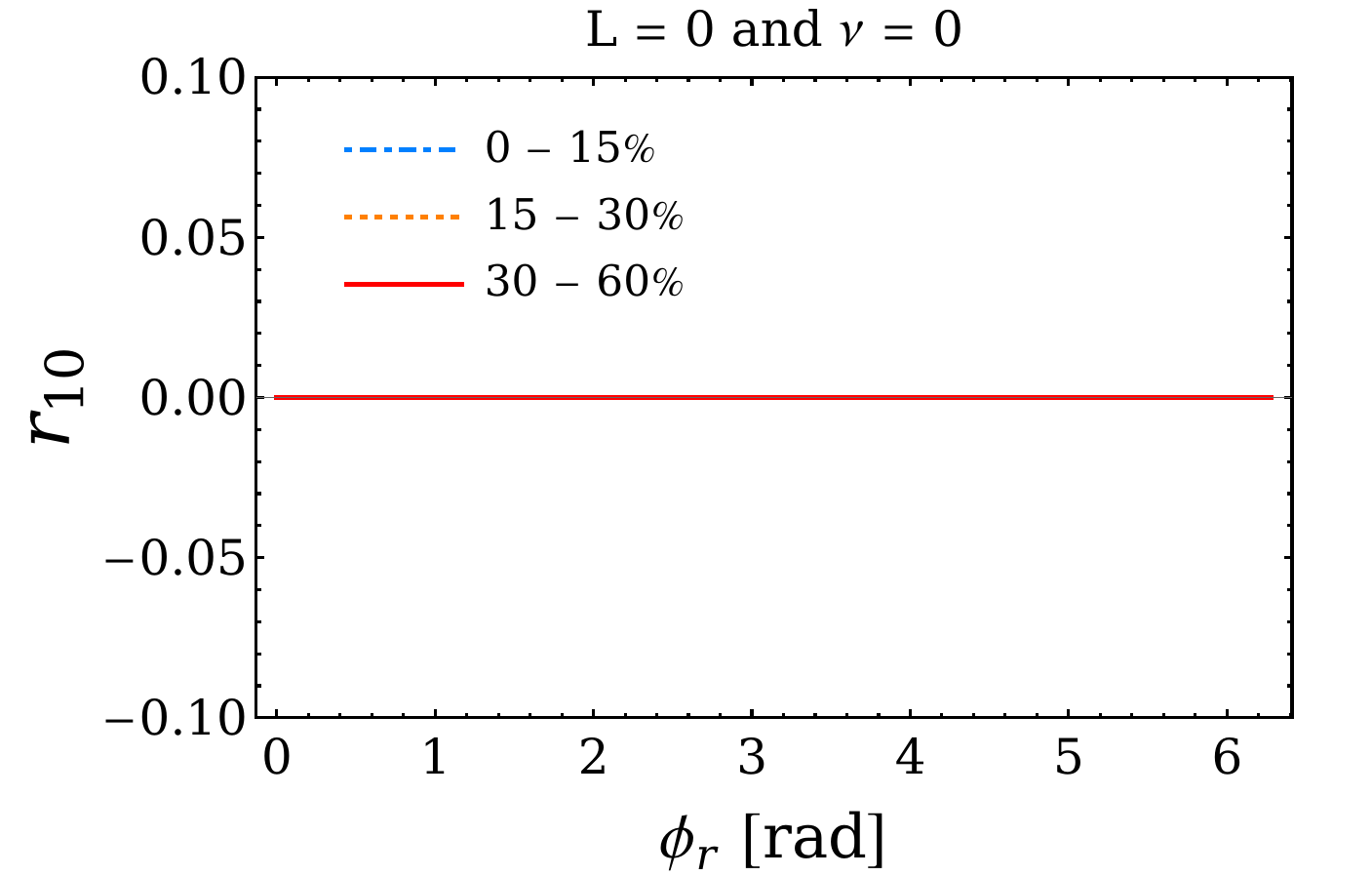}
   \epsfig{width=0.49\textwidth,figure=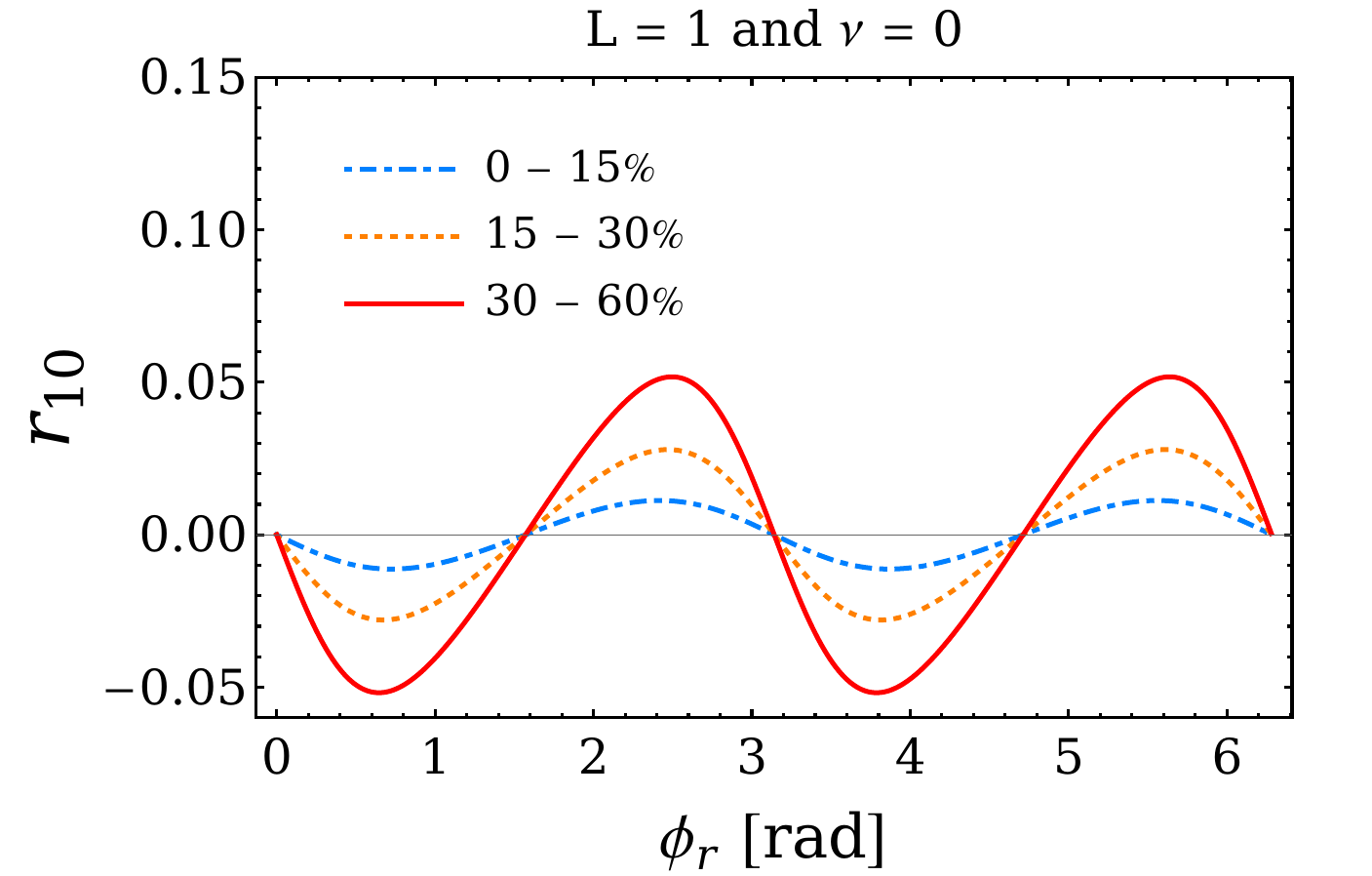}
   \epsfig{width=0.49\textwidth,figure=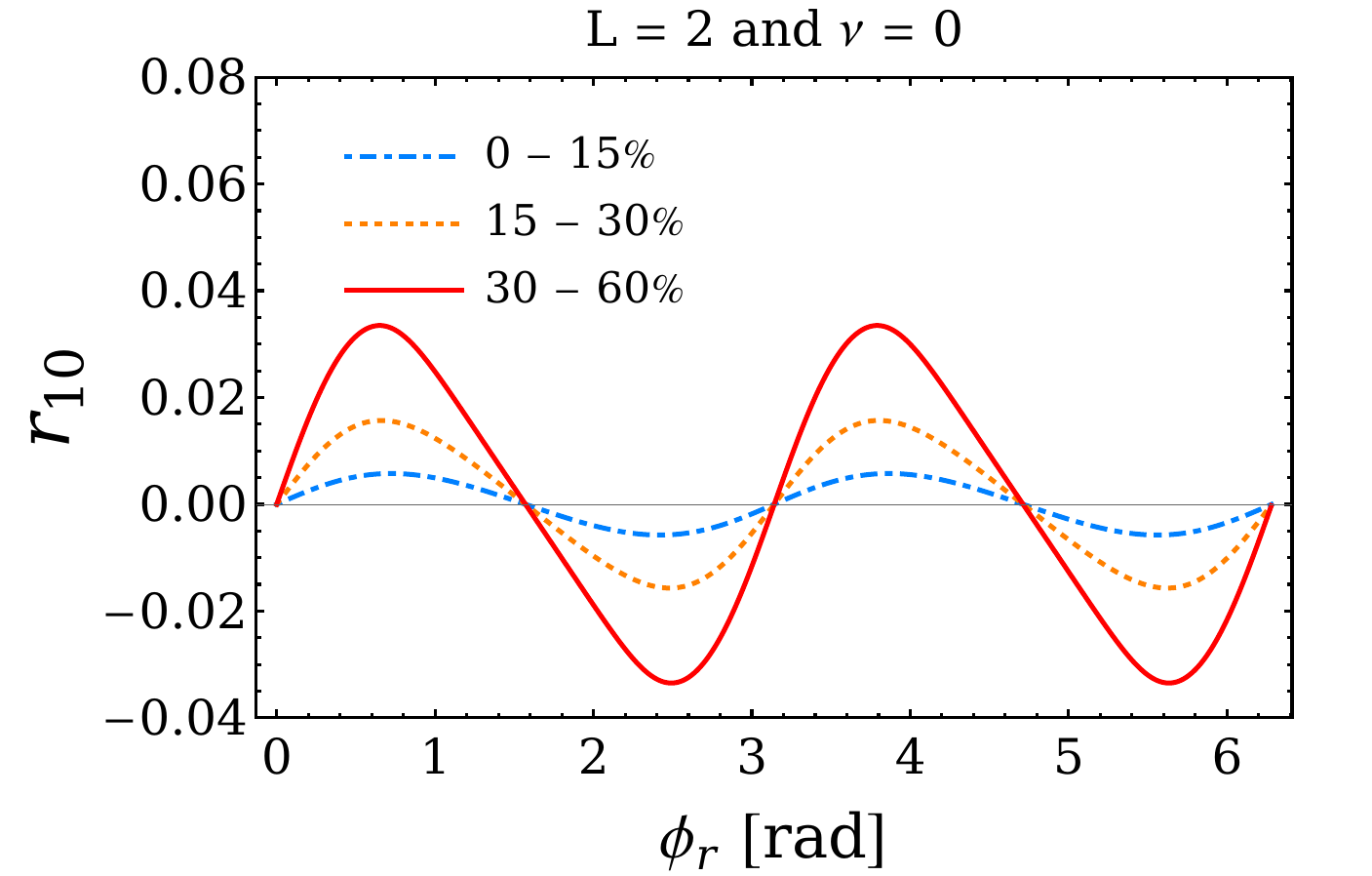}
    \epsfig{width=0.49\textwidth,figure=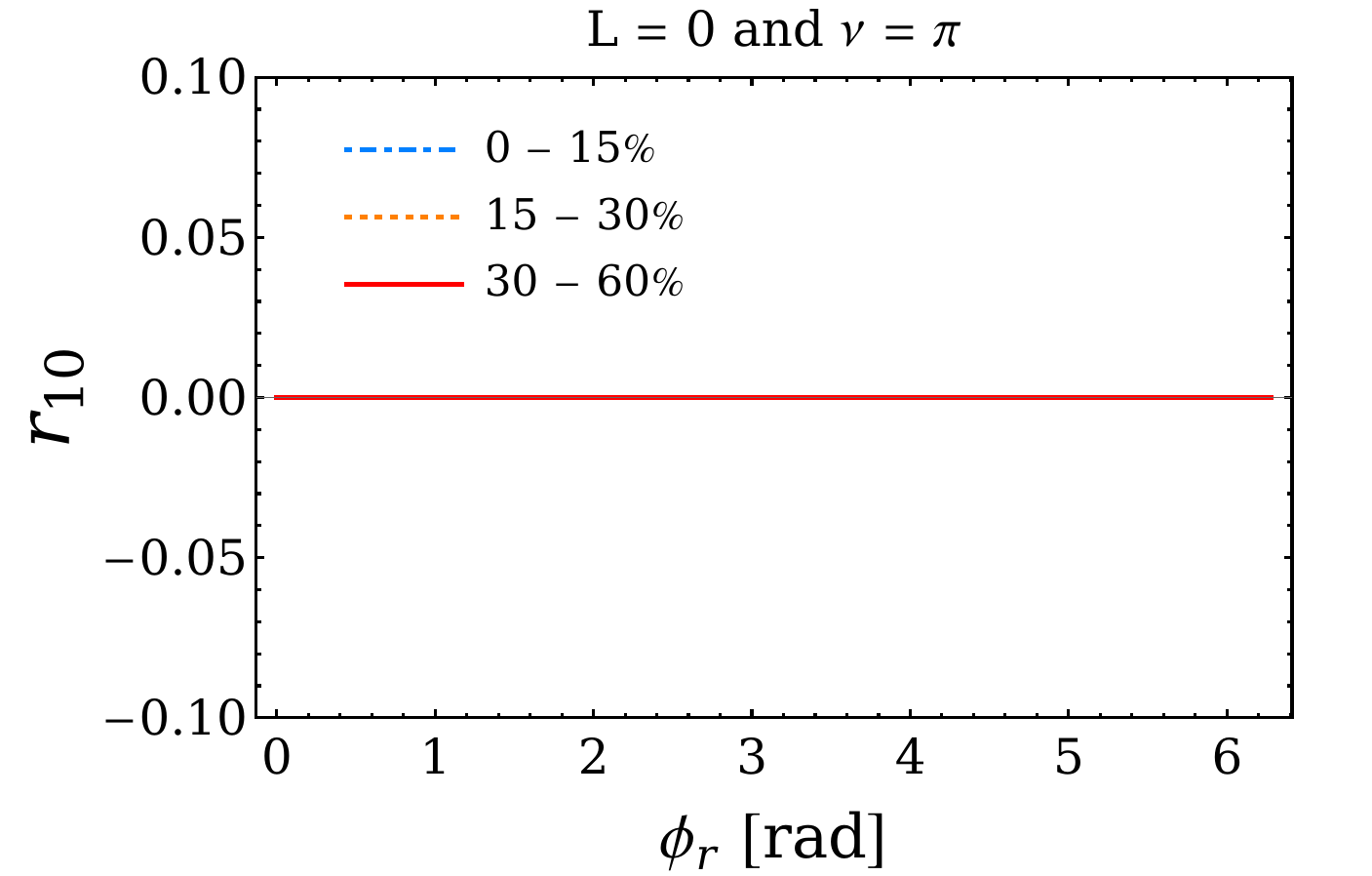}
   \epsfig{width=0.49\textwidth,figure=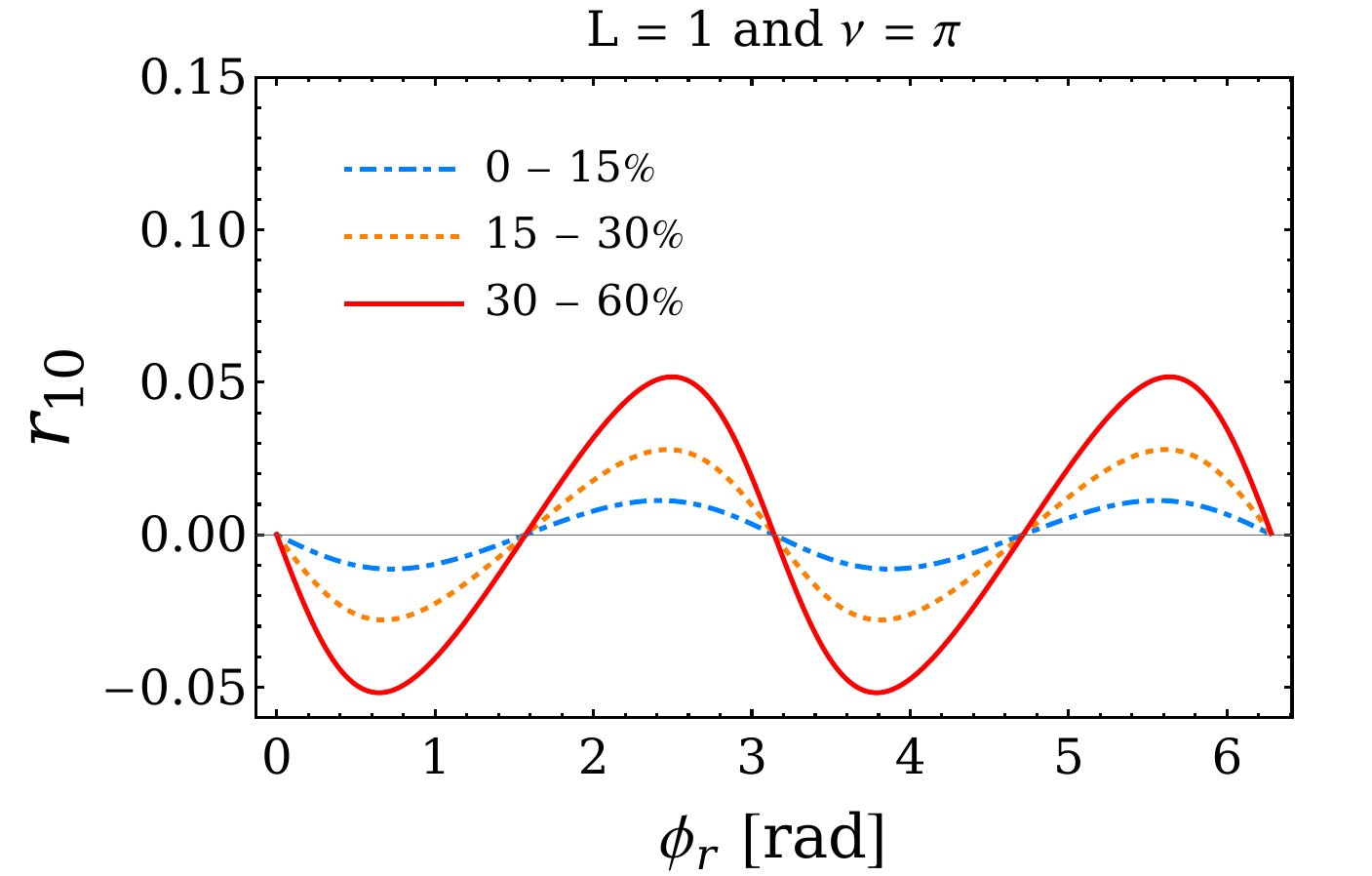}
   \epsfig{width=0.49\textwidth,figure=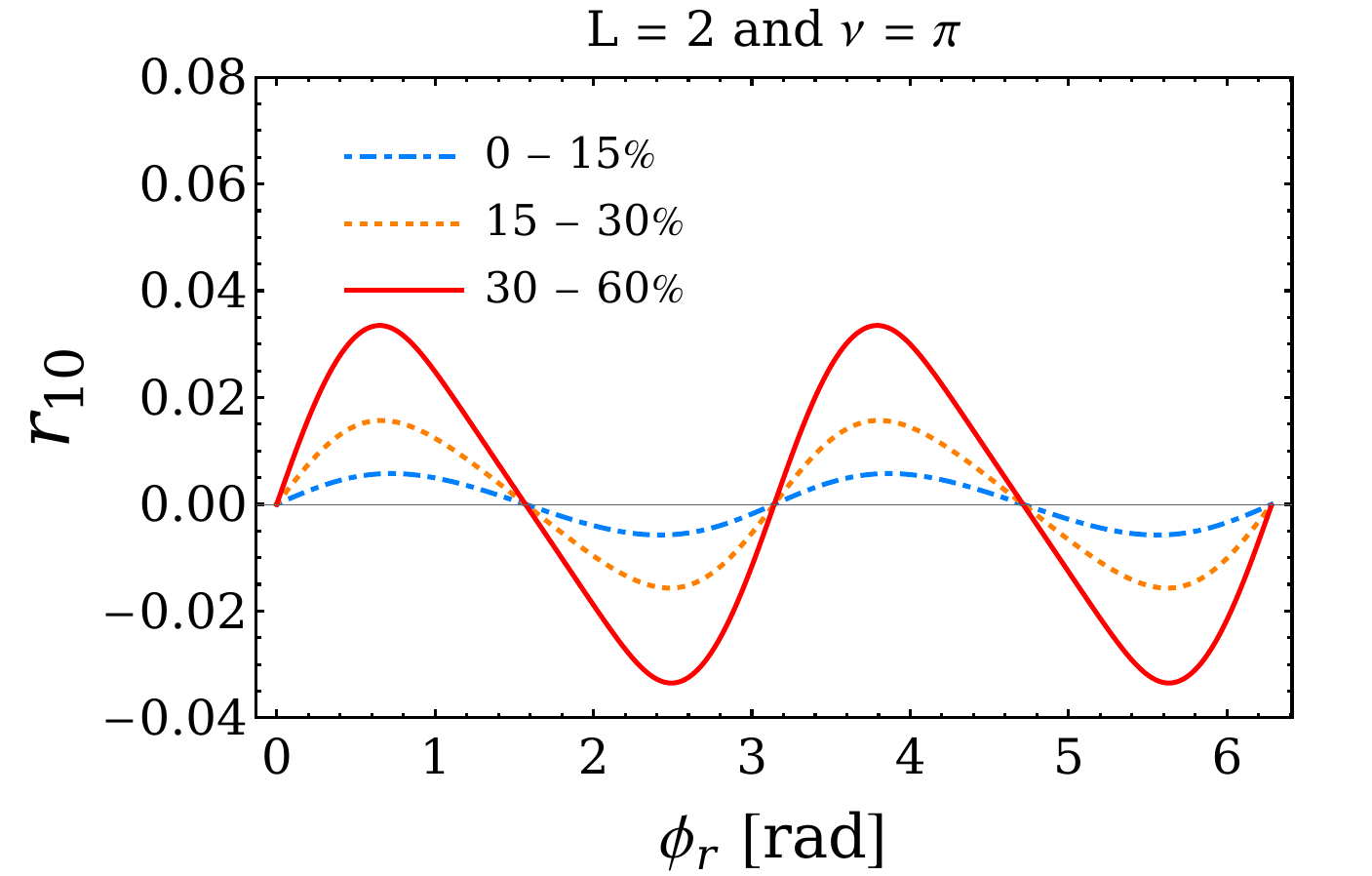}
            \caption{The density matrix coefficient $r_{10}$ with respect to the reaction plane azimuthal angle $\phi_r$ for the angles between spin and vorticity $\nu = 0$ and $\nu = \pi$. \label{r10plot}}
        \end{center}
\end{figure}

\begin{figure}[h]
		\begin{center}
    \epsfig{width=0.49\textwidth,figure=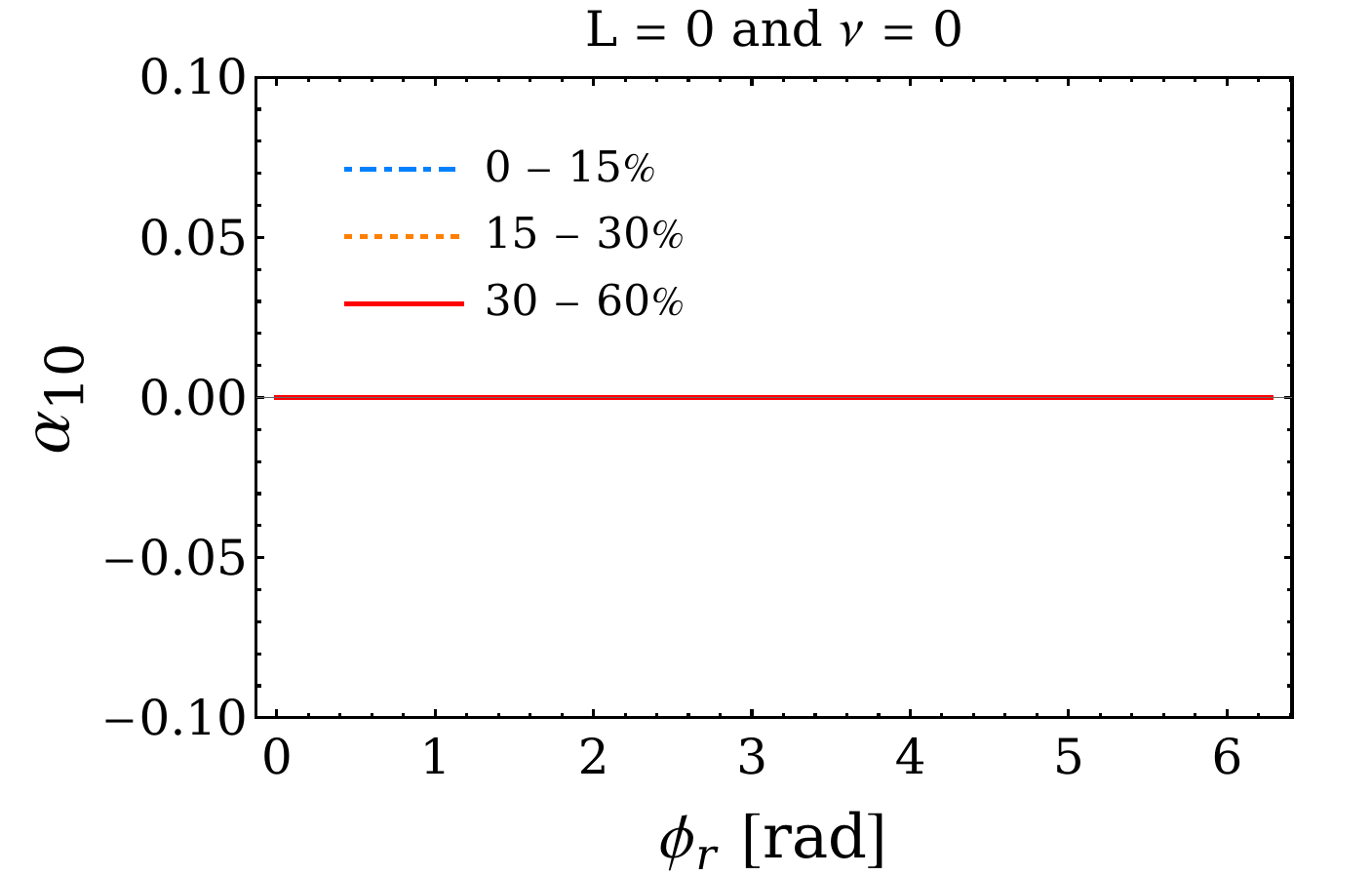}
   \epsfig{width=0.49\textwidth,figure=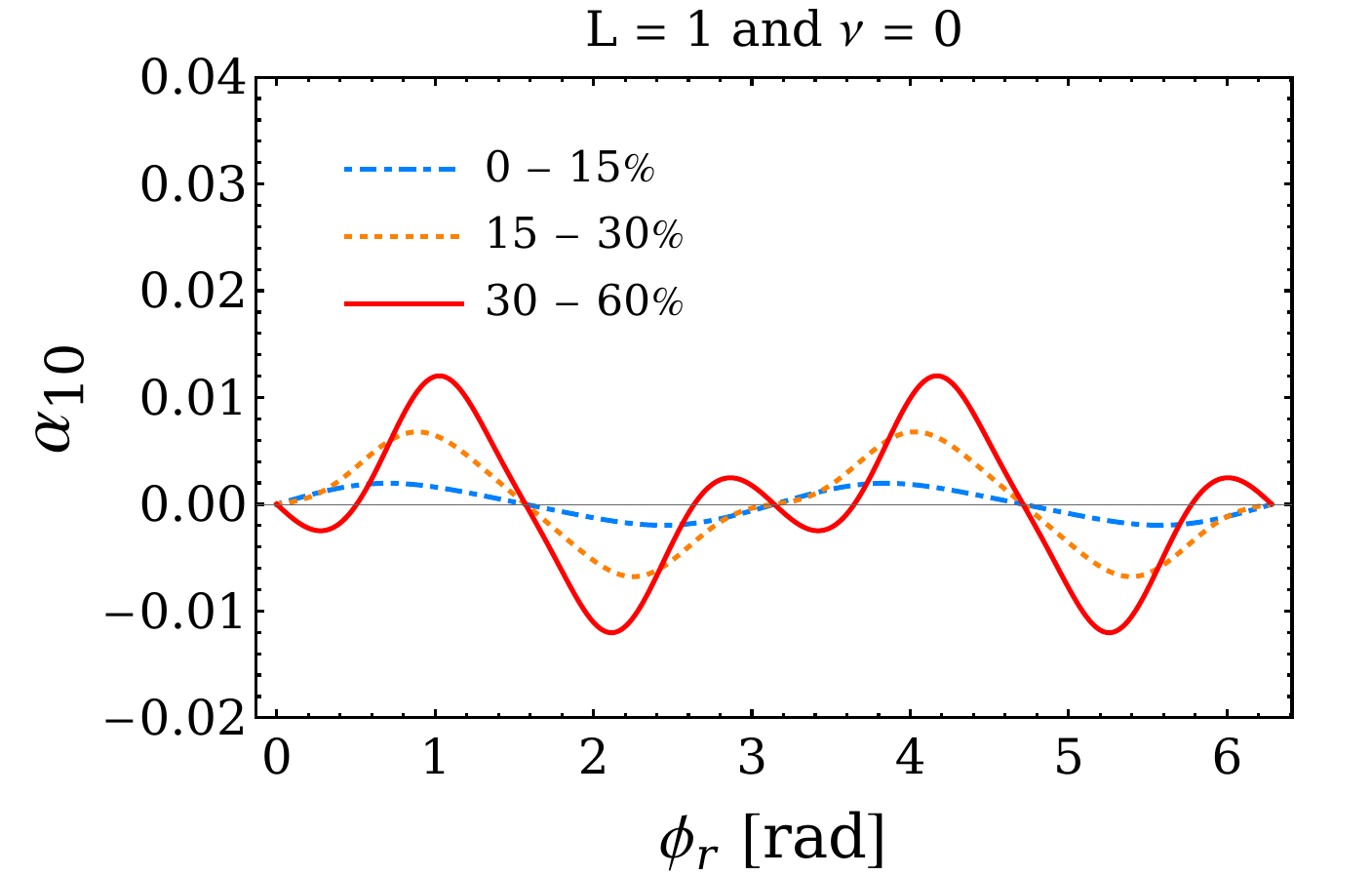}
   \epsfig{width=0.49\textwidth,figure=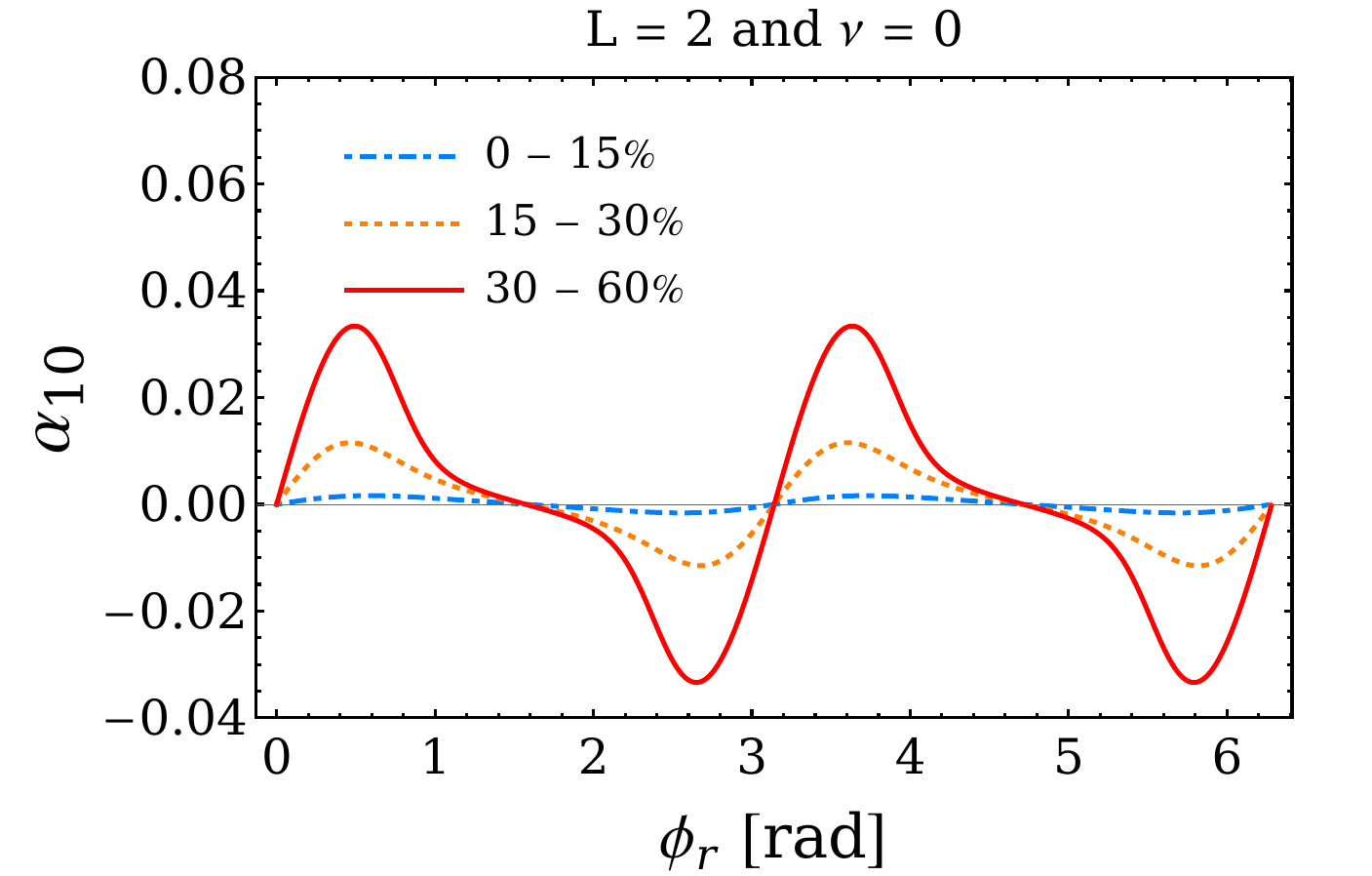}
    \epsfig{width=0.49\textwidth,figure=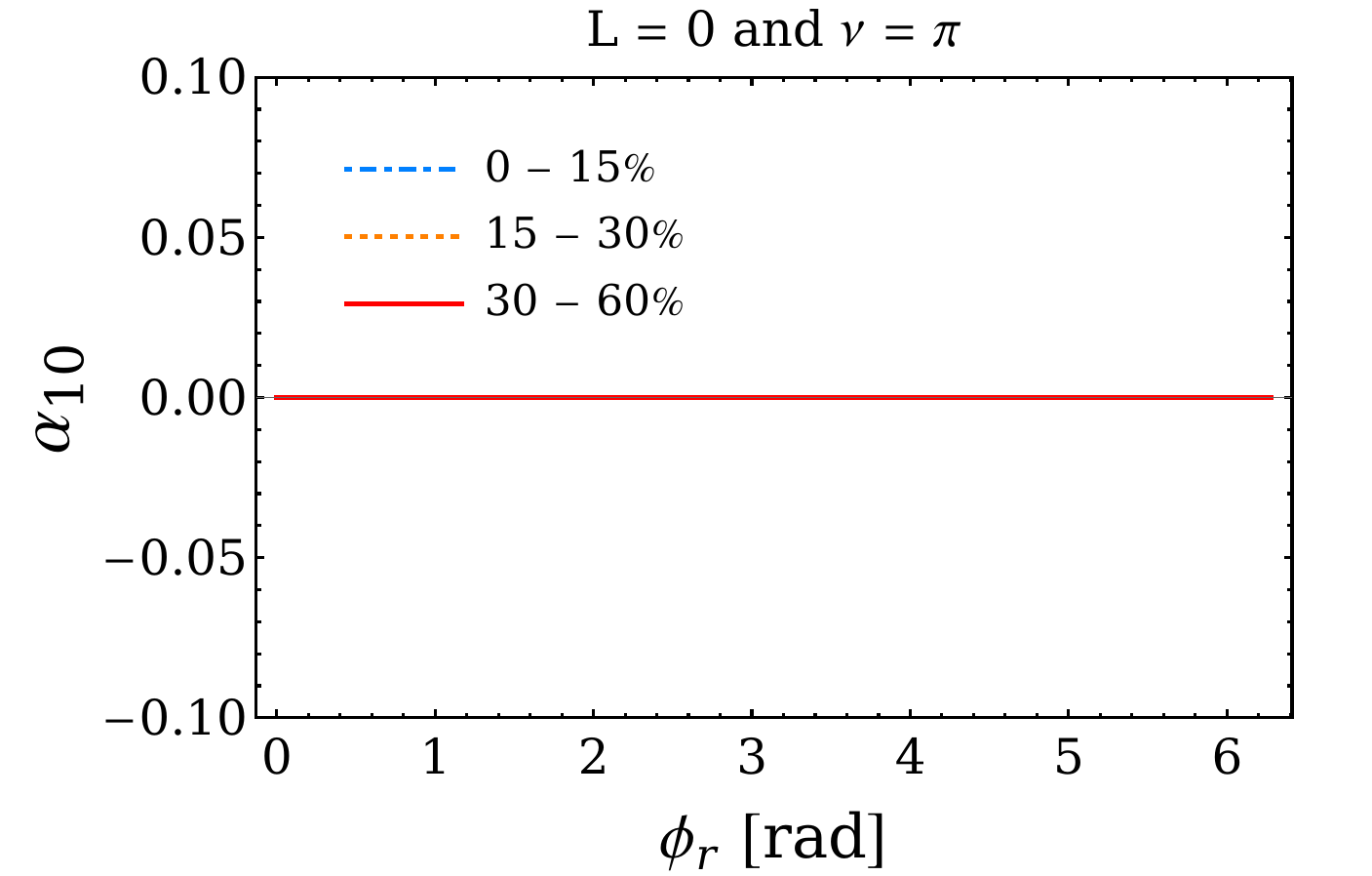}
   \epsfig{width=0.49\textwidth,figure=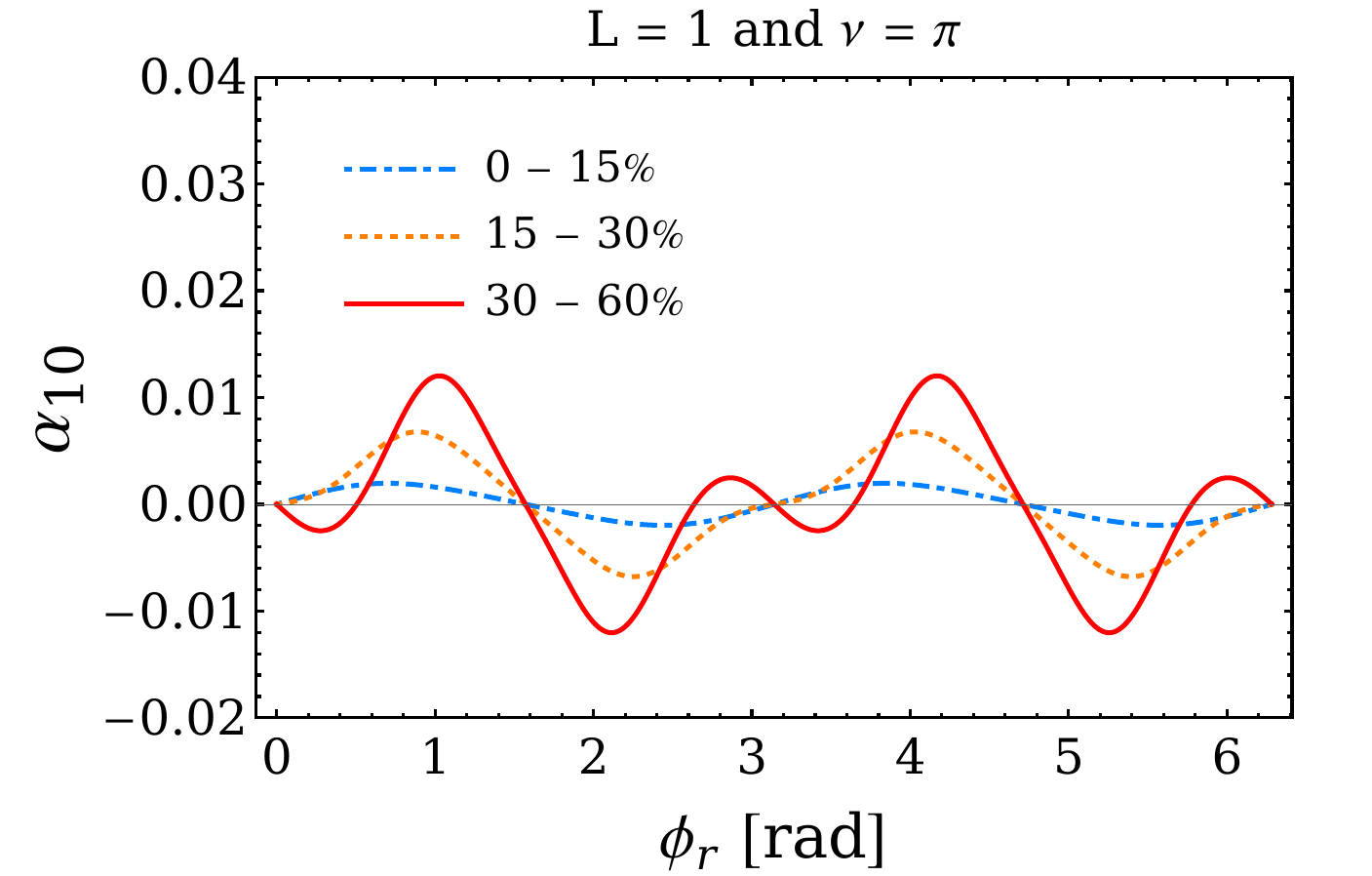}
   \epsfig{width=0.49\textwidth,figure=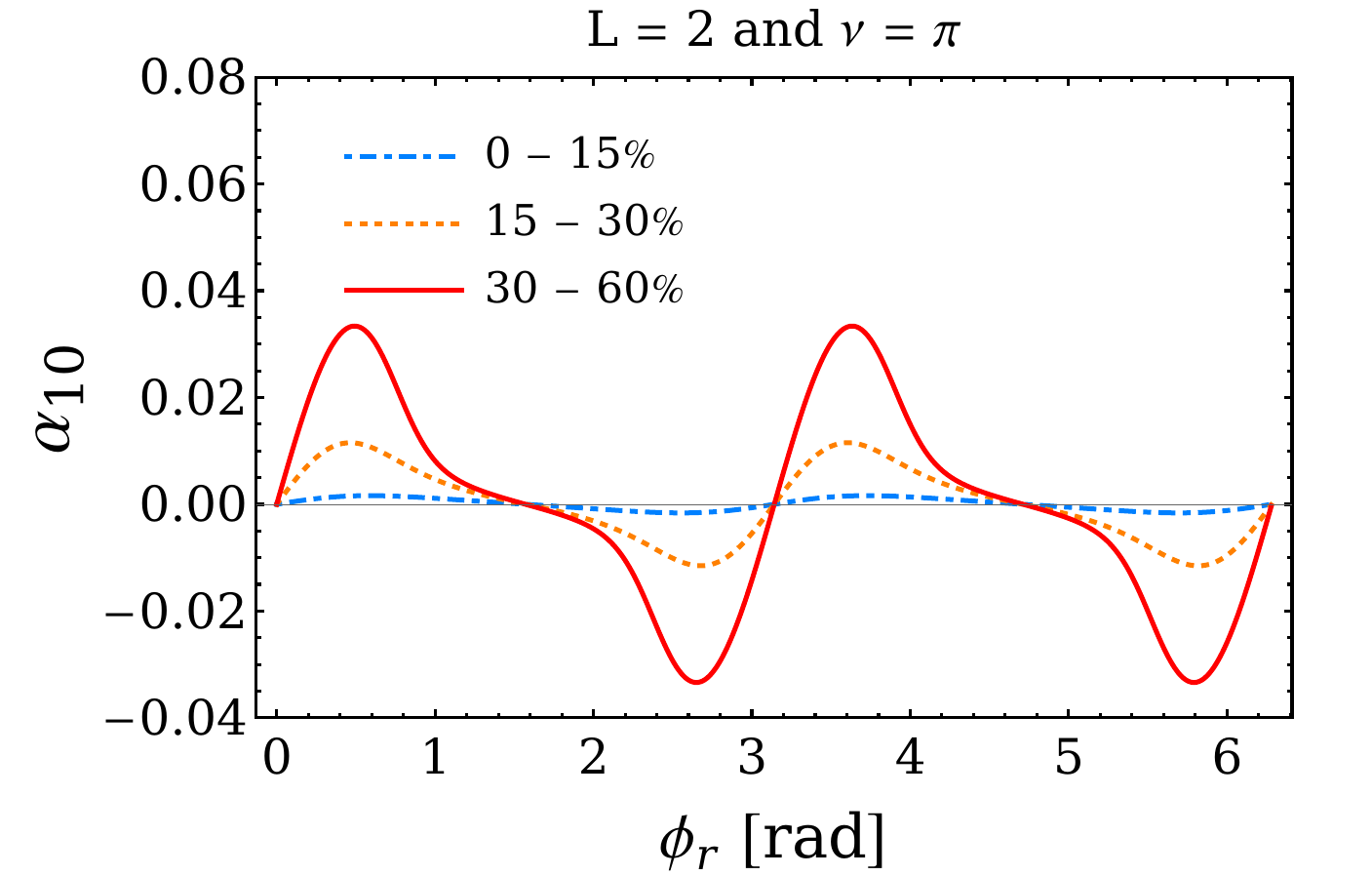}
            \caption[]{The off-diagonal density matrix coefficient $\alpha_{10}$ in relation to azimuthal angle $\phi_r$ for the angle between spin and vorticity $\nu = 0$ and $\nu = \pi$.\label{alpha10plot}}
        \end{center}
\end{figure}

\begin{figure}[h]
		\begin{center}
    \epsfig{width=0.49\textwidth,figure=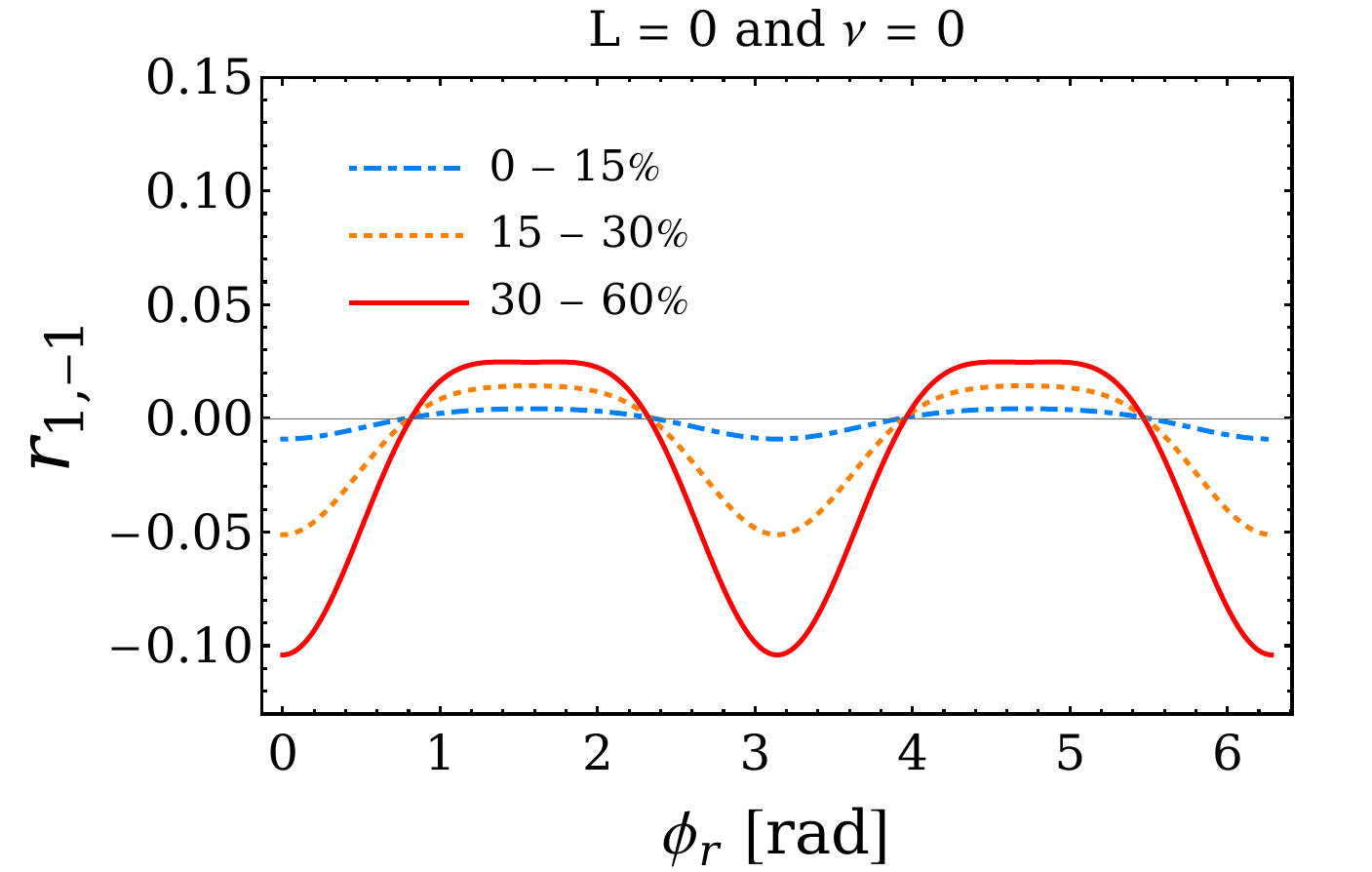}
   \epsfig{width=0.49\textwidth,figure=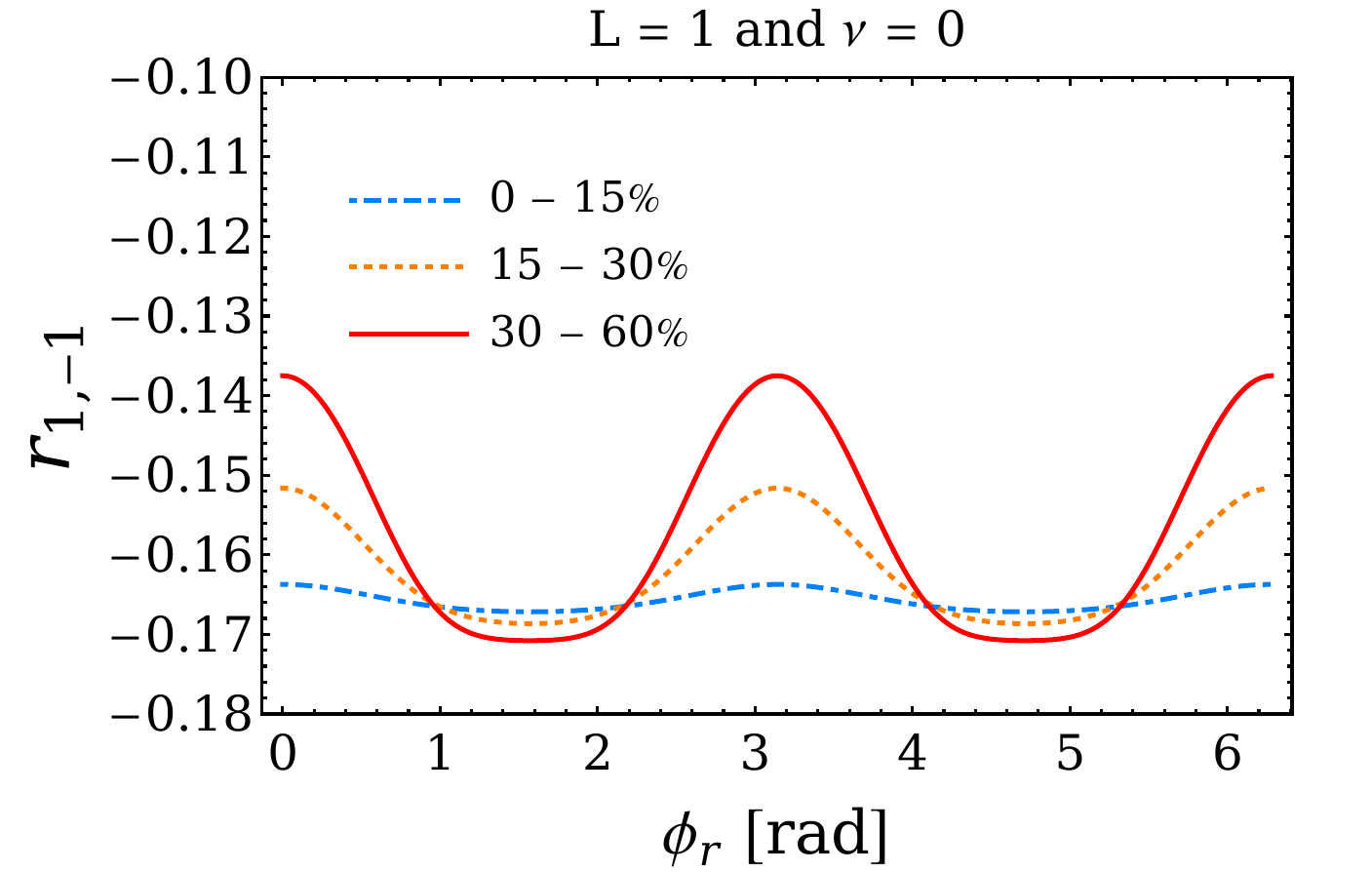}
   \epsfig{width=0.49\textwidth,figure=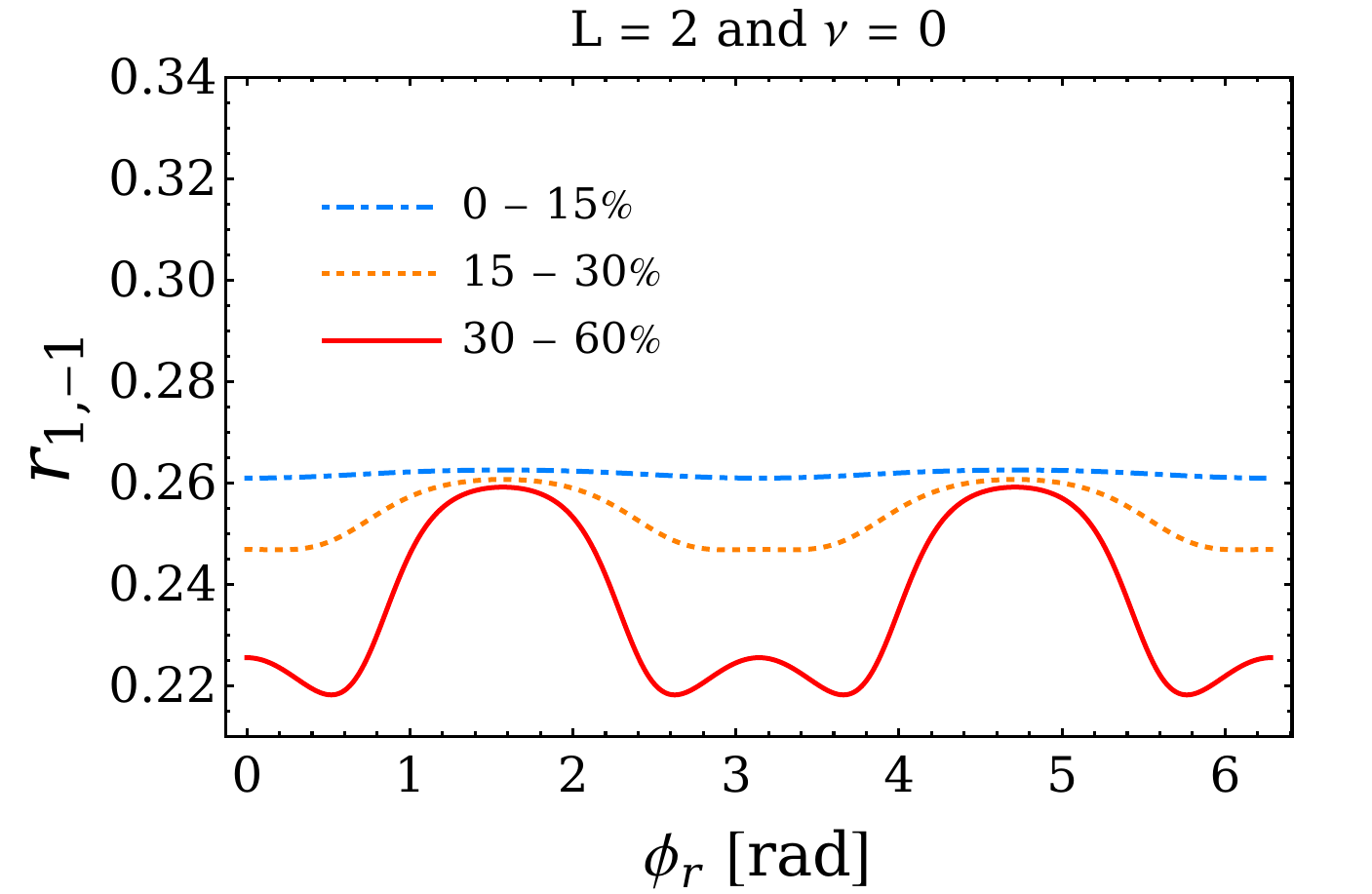}
   \epsfig{width=0.49\textwidth,figure=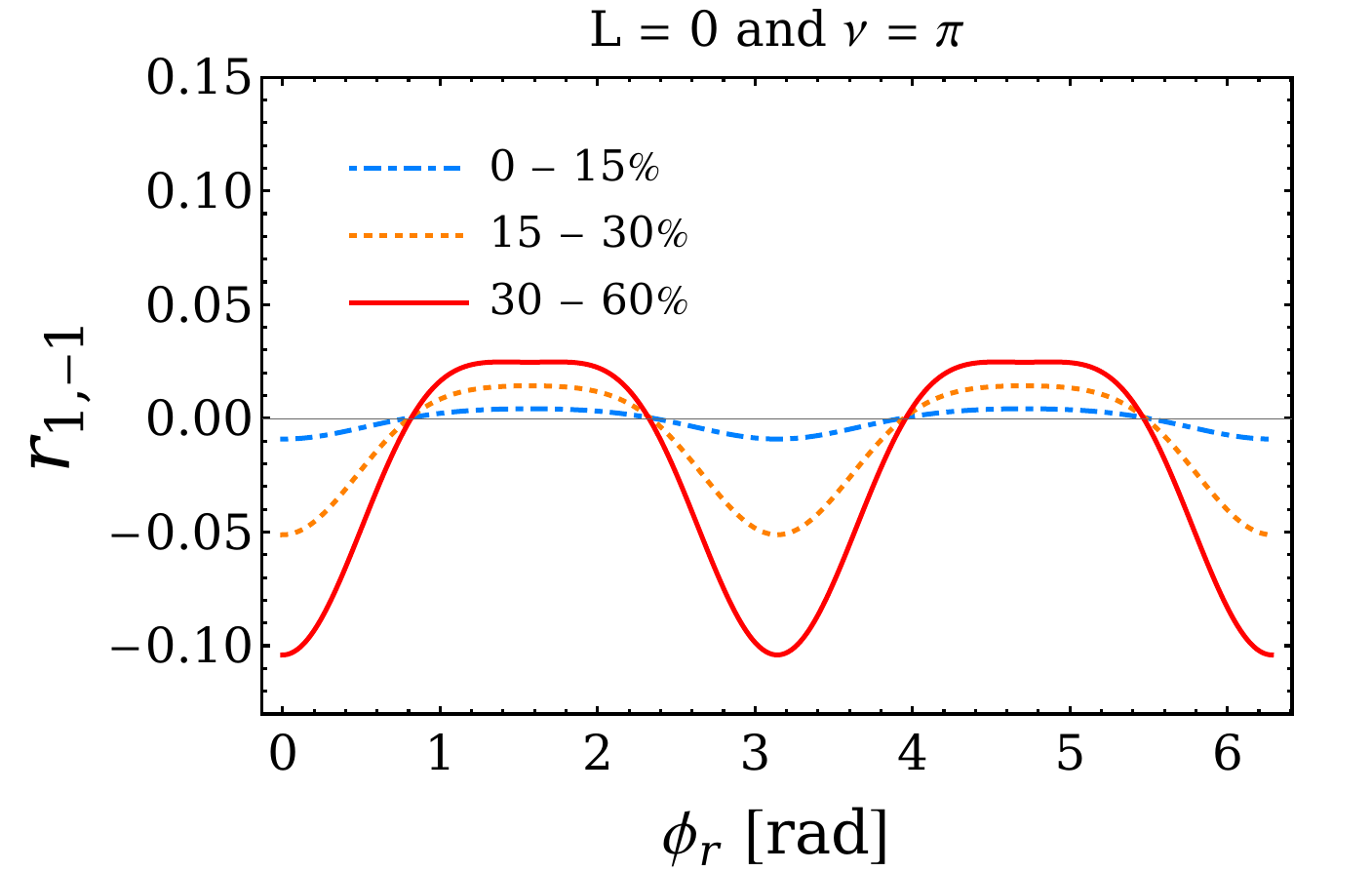}
   \epsfig{width=0.49\textwidth,figure=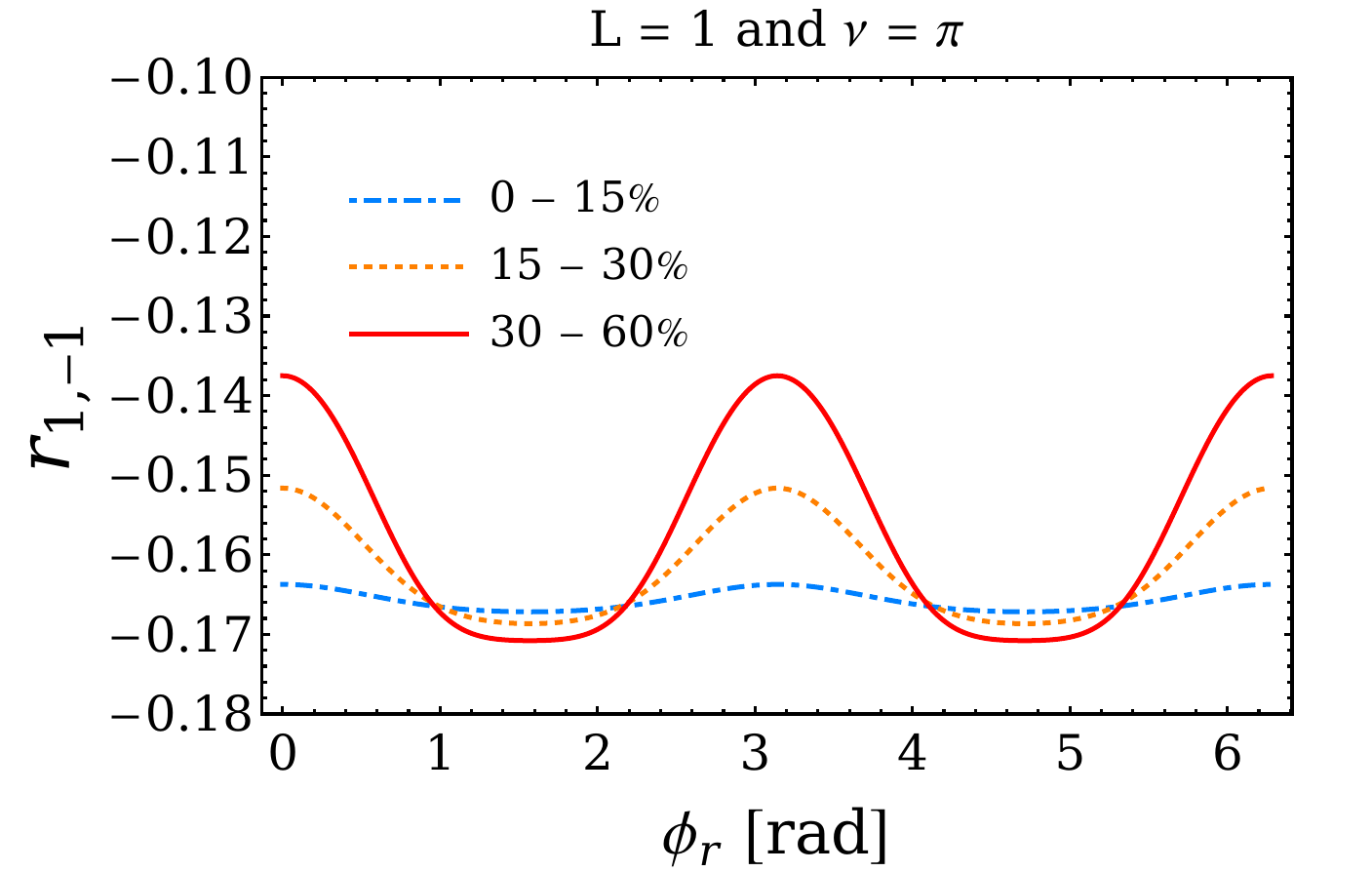}
   \epsfig{width=0.49\textwidth,figure=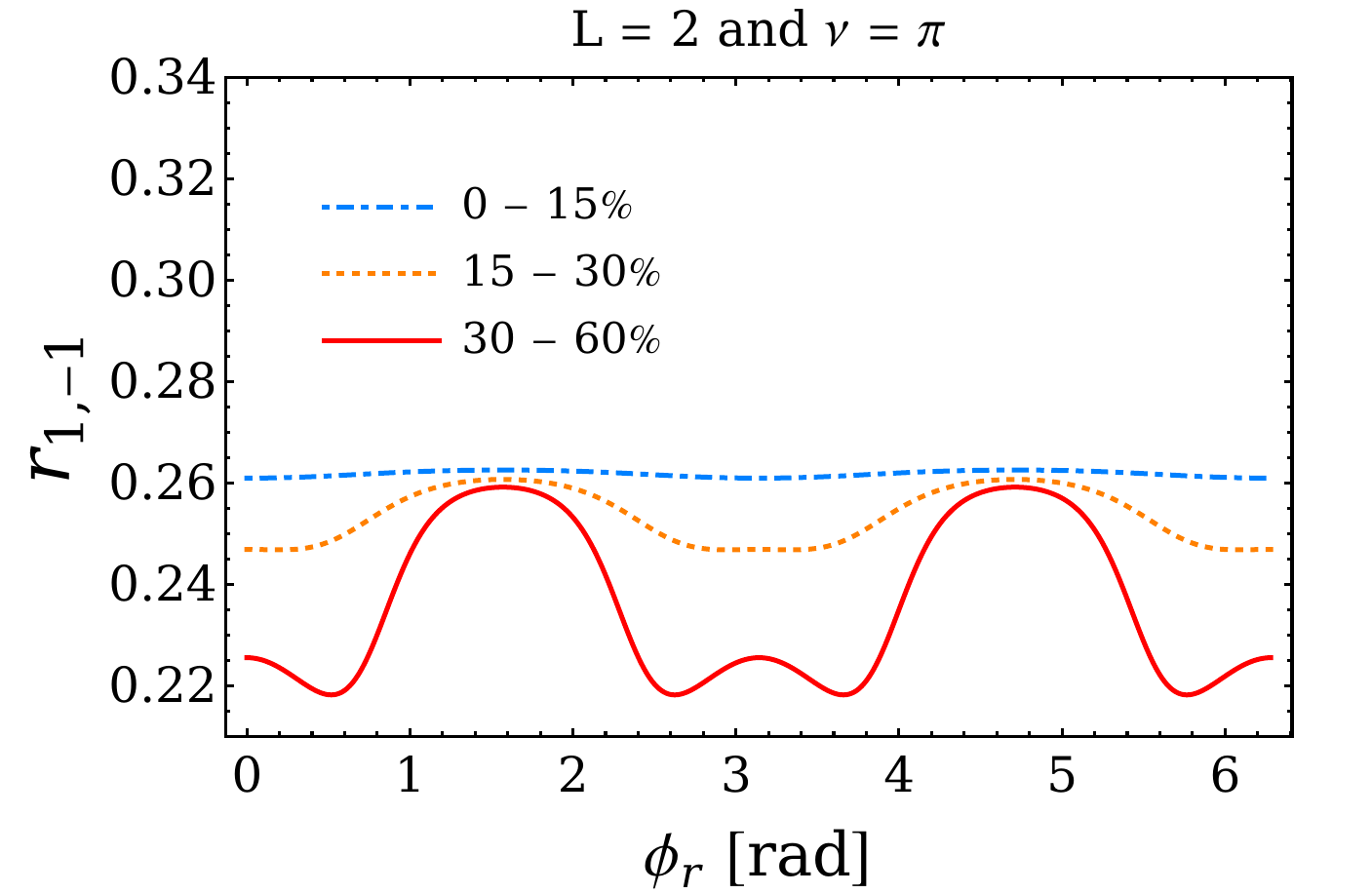}
            \caption[]{The off-diagonal density matrix coefficient $r_{1,-1}$ for the case where the angle between the spin and vorticity $\nu_0=0$ and $\nu=\pi$, in relation to the azimuthal angle $\phi_r$ for various centralities. \label{rpmplus}}
        \end{center}
\end{figure}
\begin{figure}[h]
		\begin{center}
    \epsfig{width=0.49\textwidth,figure=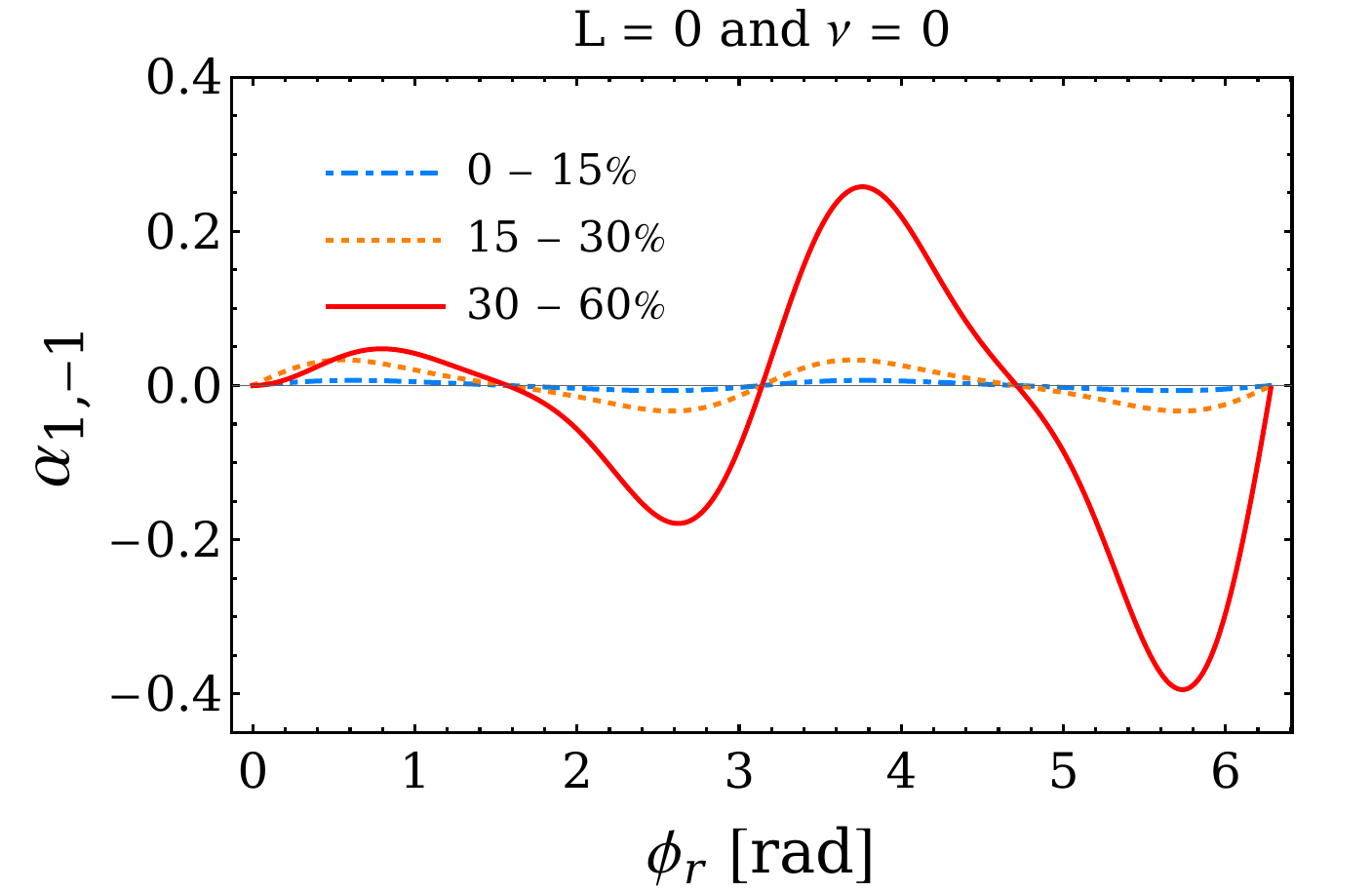}
   \epsfig{width=0.49\textwidth,figure=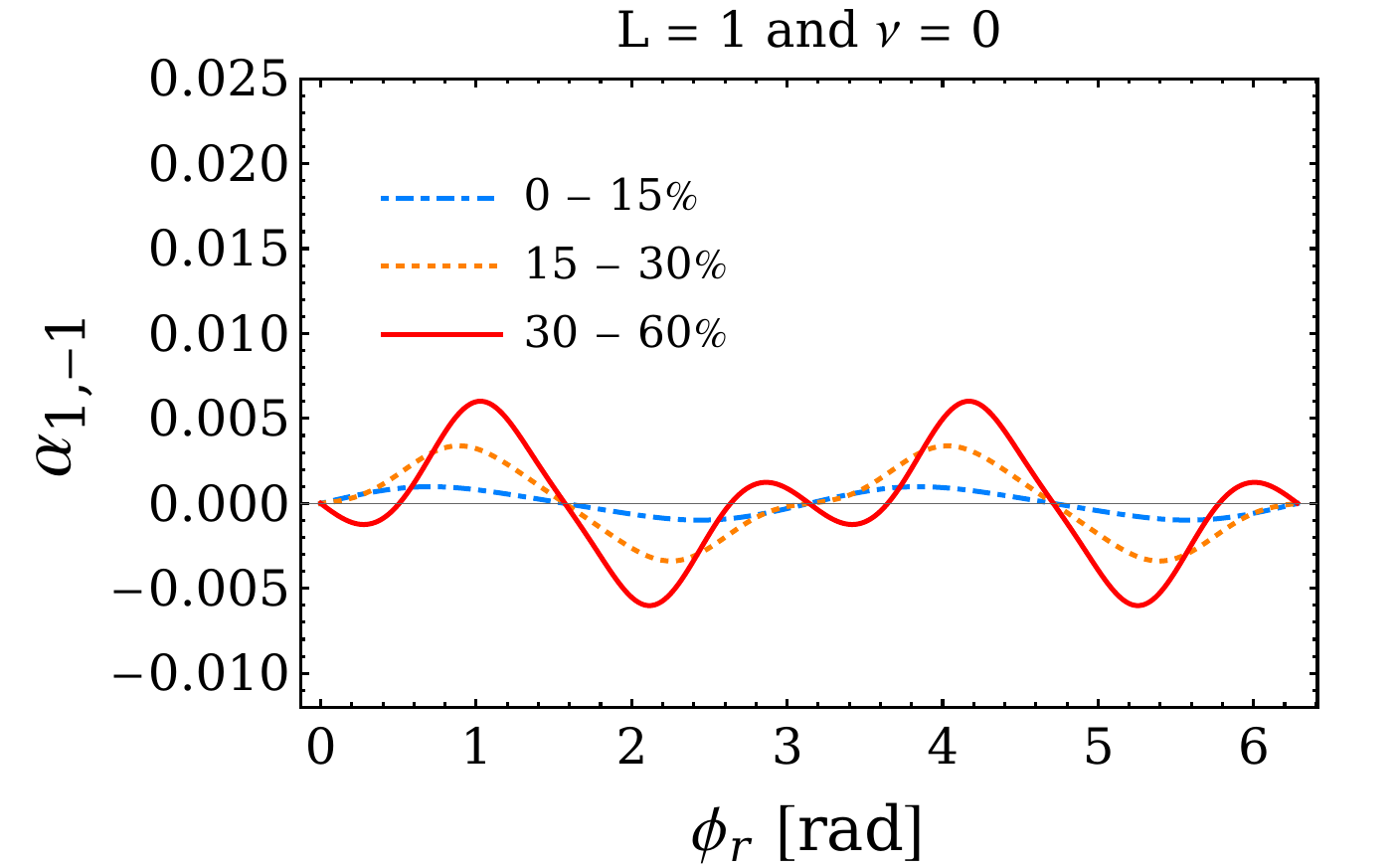}
   \epsfig{width=0.49\textwidth,figure=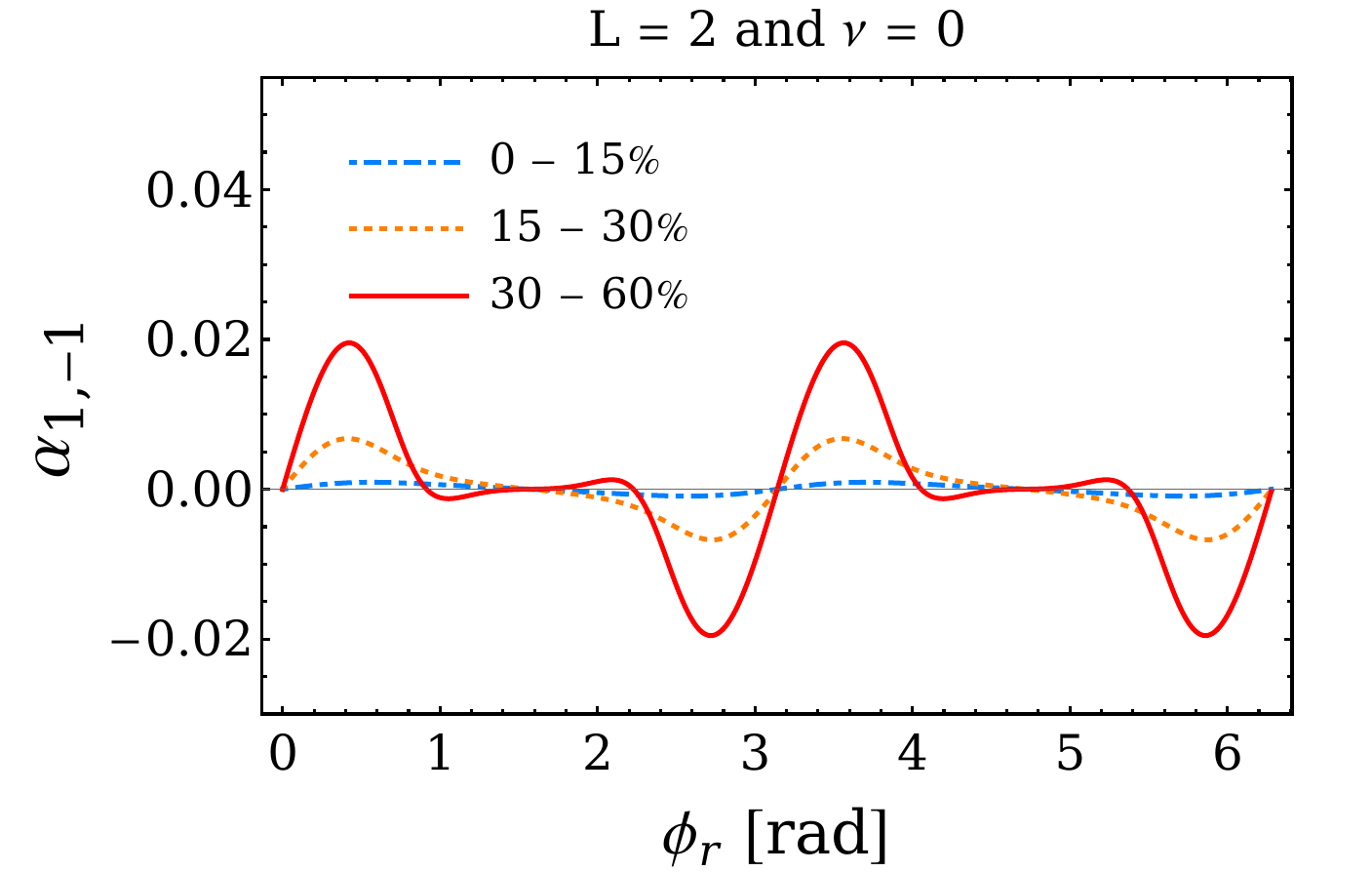}
   \epsfig{width=0.49\textwidth,figure=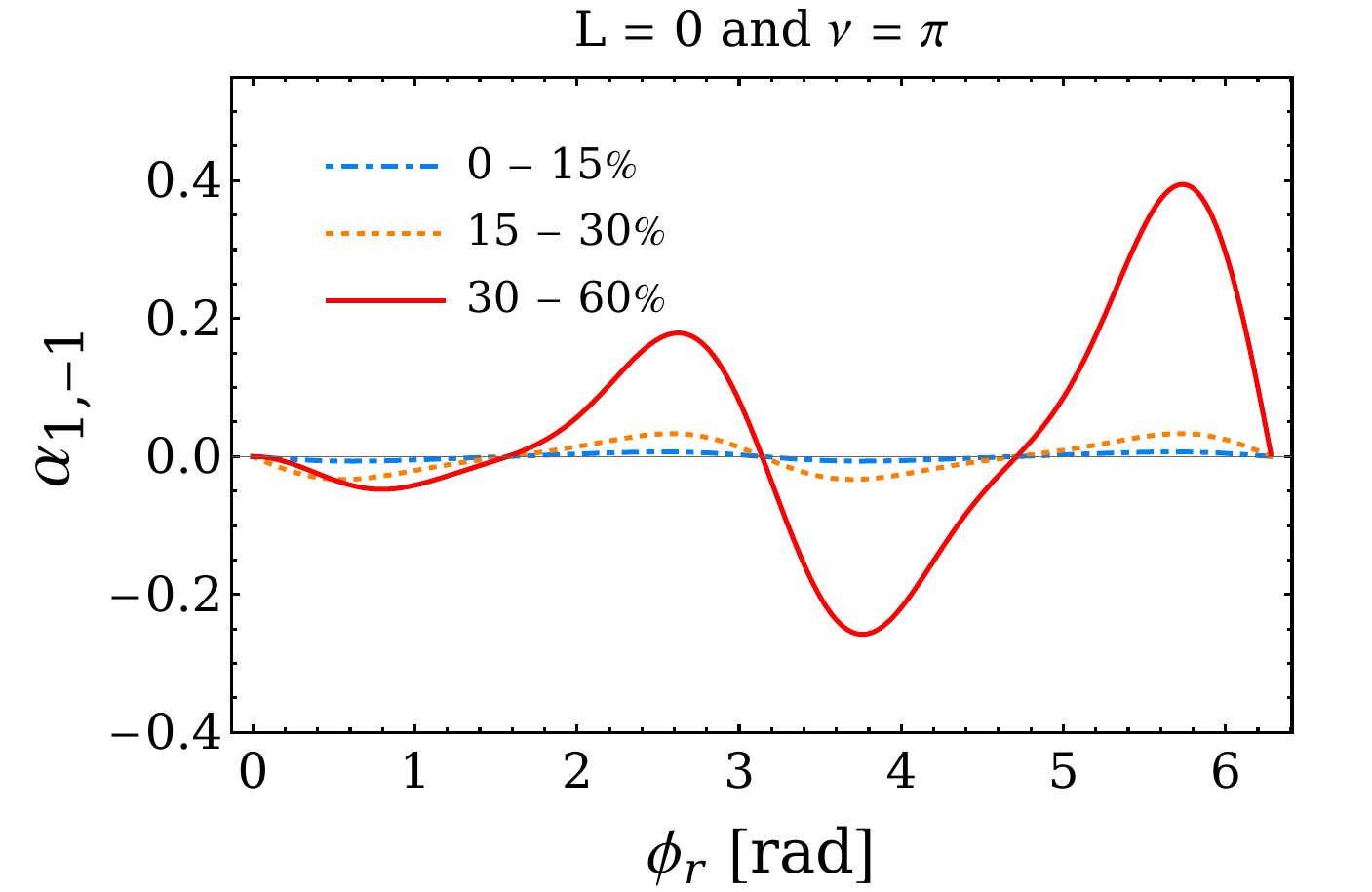}
   \epsfig{width=0.49\textwidth,figure=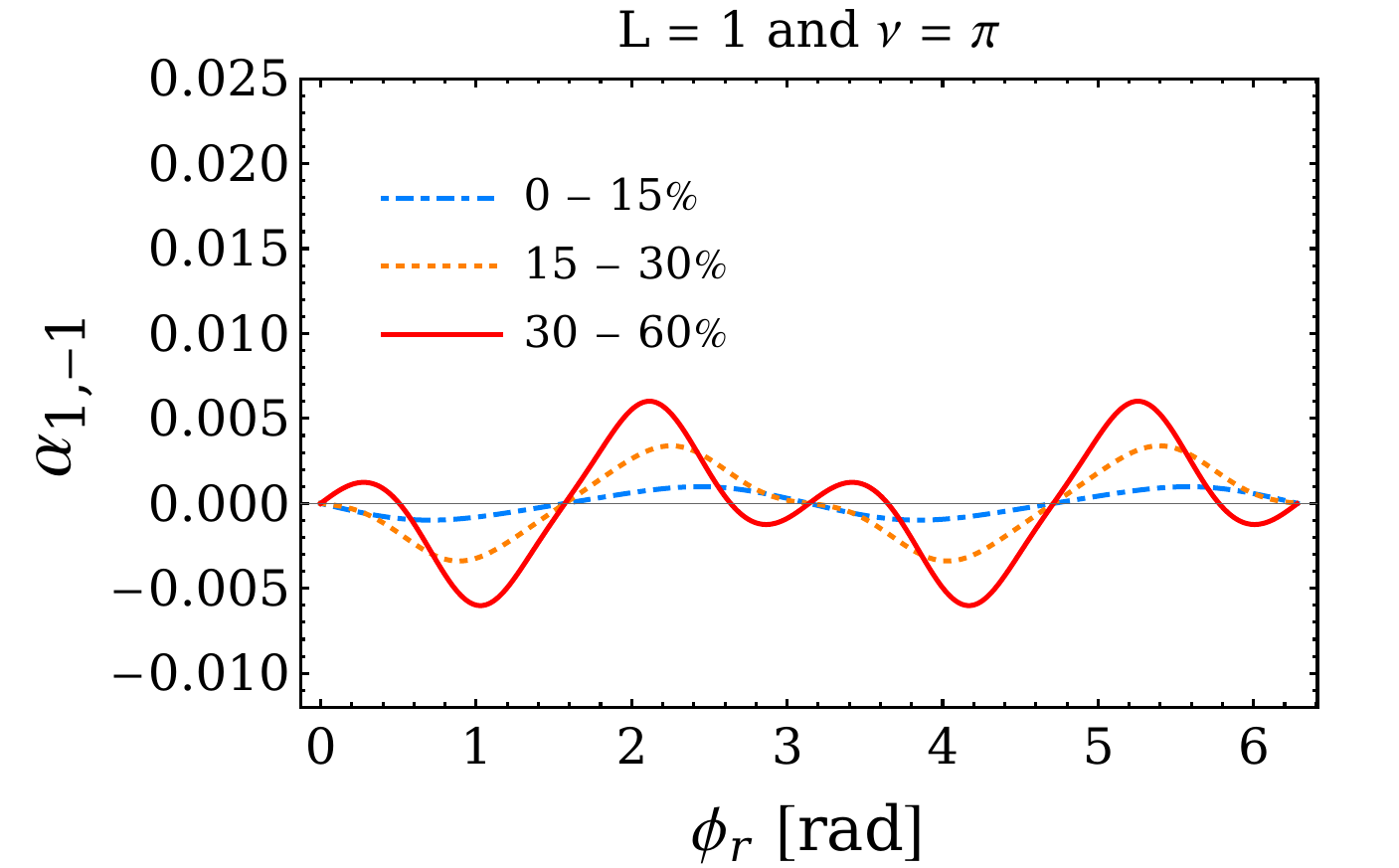}
   \epsfig{width=0.49\textwidth,figure=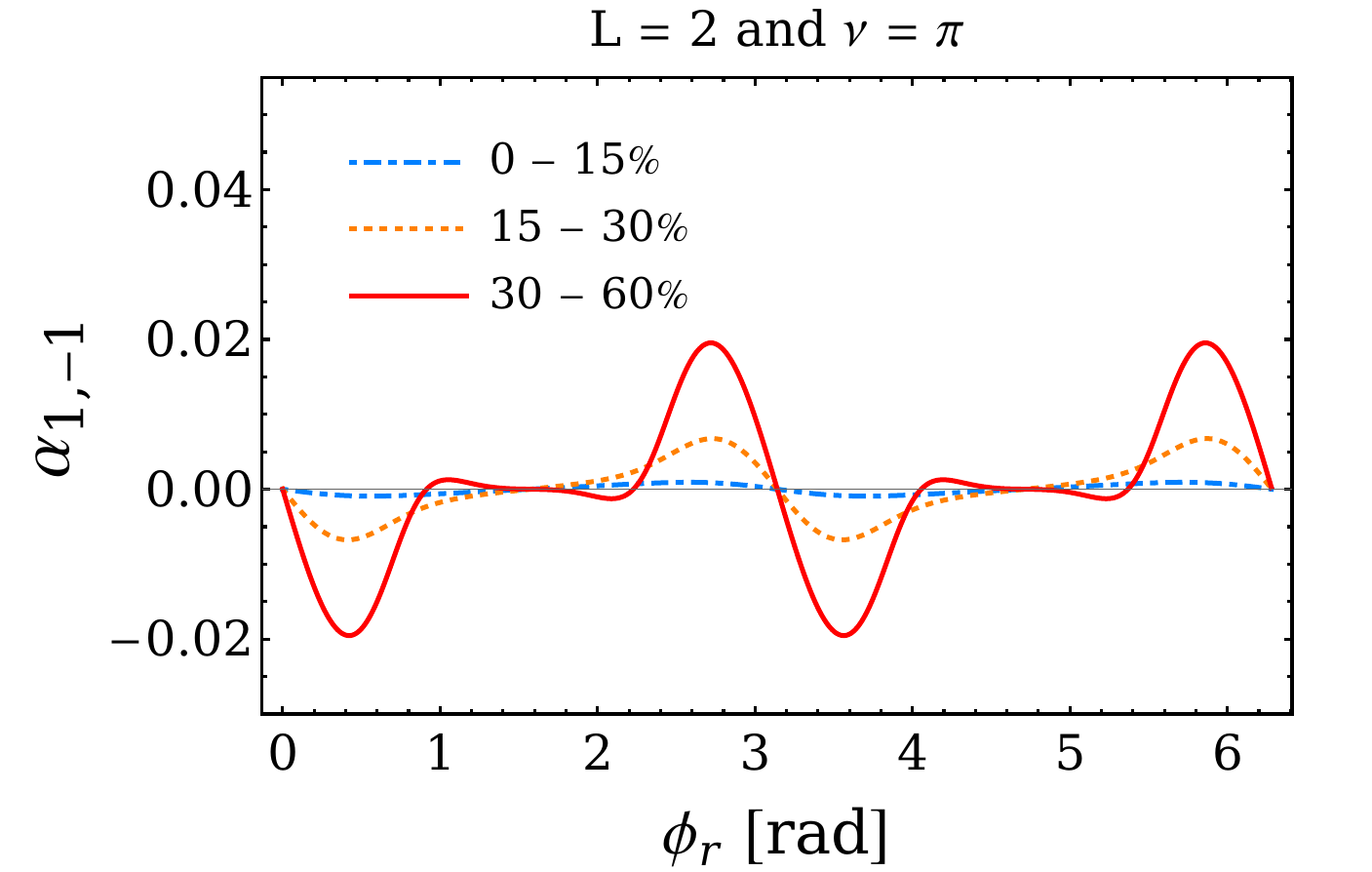}
            \caption[]{The off-diagonal density matrix coefficient $\alpha_{1,-1}$
  in relation to the azimuthal angle $\phi_r$ is given for angles between spin and vorticity $\nu = 0$ and $\nu = \pi$ for various centrality values. \label{alphaplot}}
        \end{center}
\end{figure}

In Figure \ref{rhora1}, we observe the alignment factor as function of spacetime rapidity for different centralities. Here $\rho_{00}$ increases with spacetime rapidity for $L = 0$ and $L = 1$ as expected experimentally \cite{sheng}. In the case of $L=2$, however, there is a decrease in $\rho_{00}$. Since the experimental value is expected to be a mixture of these components, we anticipate a relatively small influence from this component. It is important to note that we performed this analysis for {\em spacetime} rapidity. Assuming the convolution factor in \eqref{convol} follows a Bjorken like scenario (where space-time and momentum rapidities coincide, and the parent particle rapidity distribution is approximately flat), we expect that the observed momentum rapidity will follow the same behavior.

Figures \ref{r10plot},\ref{alpha10plot} ,\ref{rpmplus} and \ref{alphaplot} show the potential of measuring the off-diagonal elements as function of the azimuthal angle $\phi_r$.
As can be seen while these elements mostly average to zero they are expected to vary considerably with the azimuthal angle in the cases that the angle between spin and vorticity is non-trivial and also a considerable amount of angular momentum is transferred from vorticity to spin.  This reinforces our previous conclusions \cite{kayman1,kayman2} that these off-diagonal matrix elements are crucial in understanding hadronization in a vortical enviroenment, for they directly probe how  different components of the angular momentum projections to the spin space density matrix.    We also notice that 
the parity w.r.t. $\phi_r$ ($\phi_r \rightarrow \phi_r+\pi$) is broken for off-diagonal observables such as $\alpha_{1,-1}$, as expected as the axes of rotation do not commute, something also present in models such as \cite{xia}.  It will be interesting therefore to see if such a breaking appears in experiment.

The only measurement so far of these observables, in quarkonium \cite{charmonium} is compatible with Cooper-Frye style maximal mixing, but as Figures \ref{r10plot},\ref{alpha10plot} ,\ref{rpmplus} and \ref{alphaplot} show it is the azimuthal dependence (not as yet measured) that is necessary to establish this.

\begin{figure}[h]
    \epsfig{width=0.49\textwidth,figure=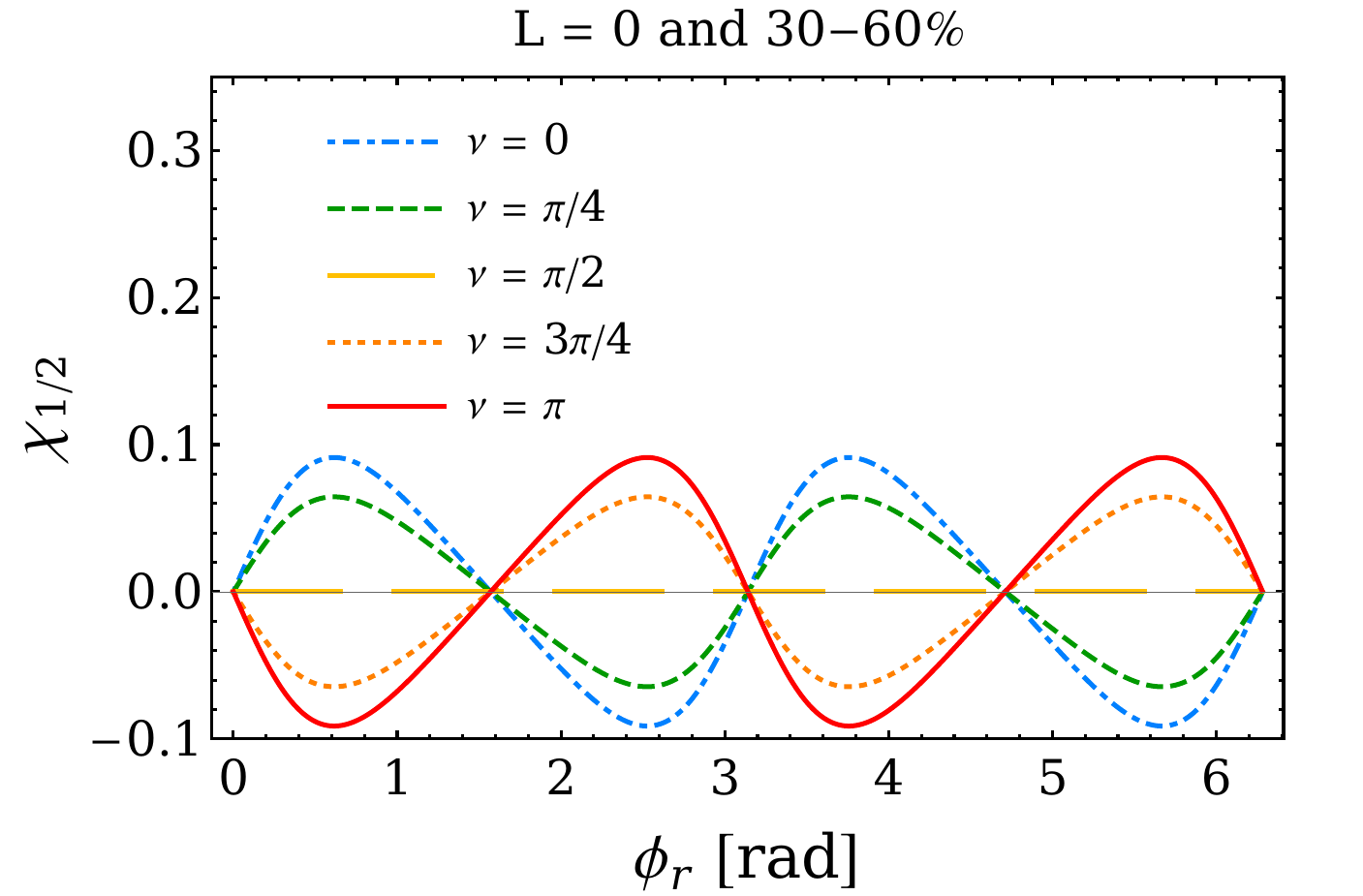}
   \epsfig{width=0.49\textwidth,figure=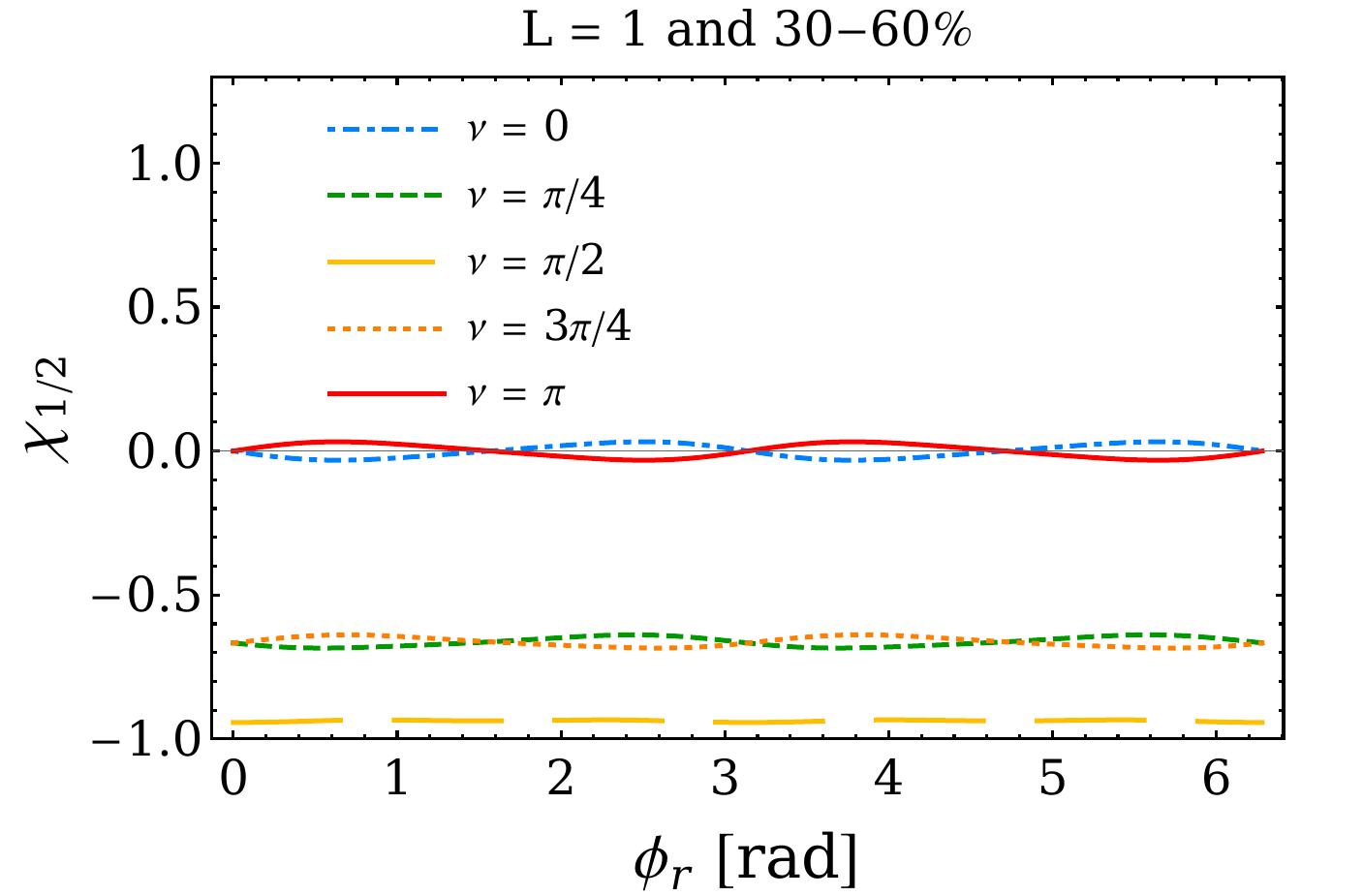}
   \epsfig{width=0.49\textwidth,figure=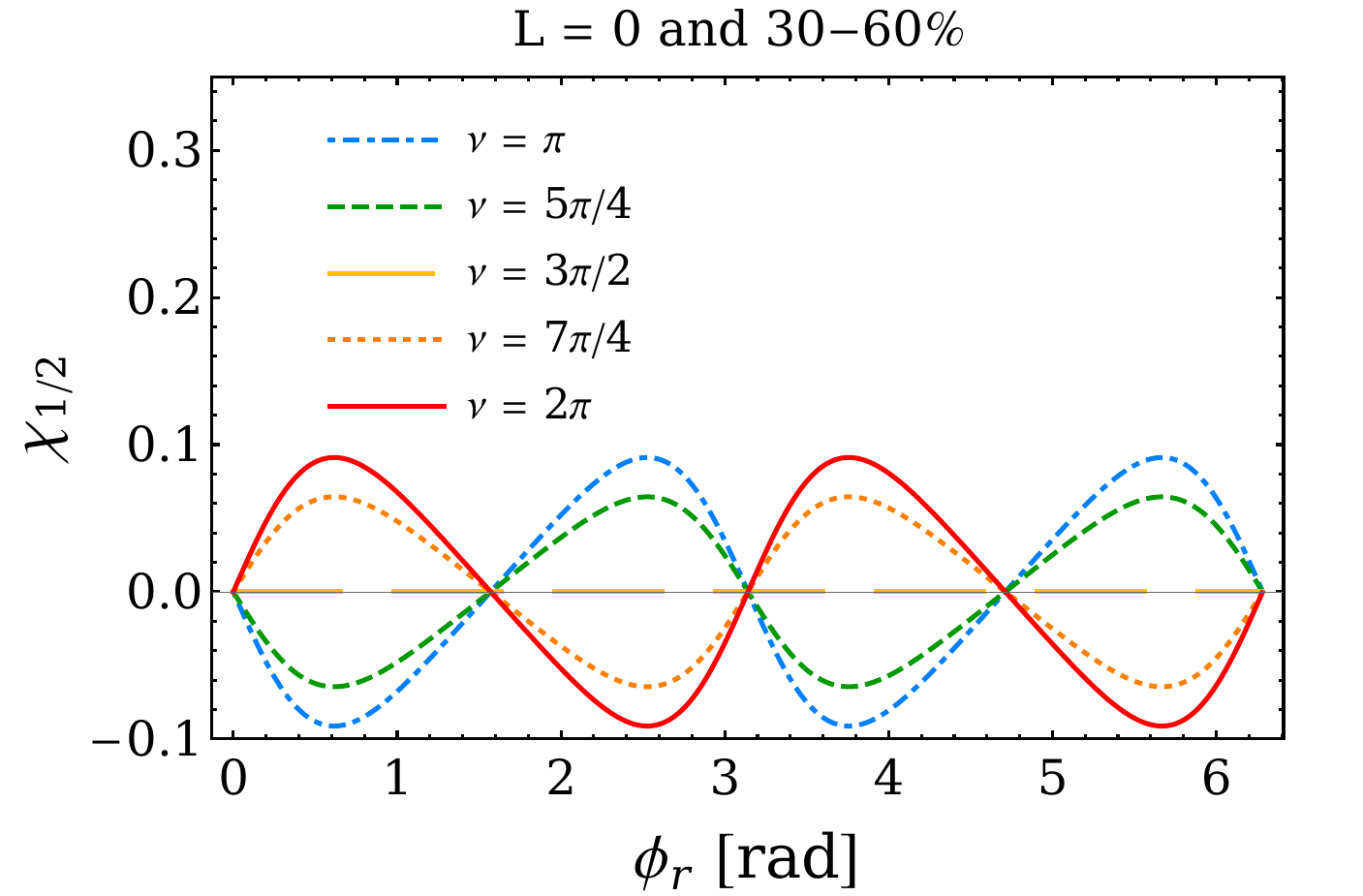}
   \epsfig{width=0.49\textwidth,figure=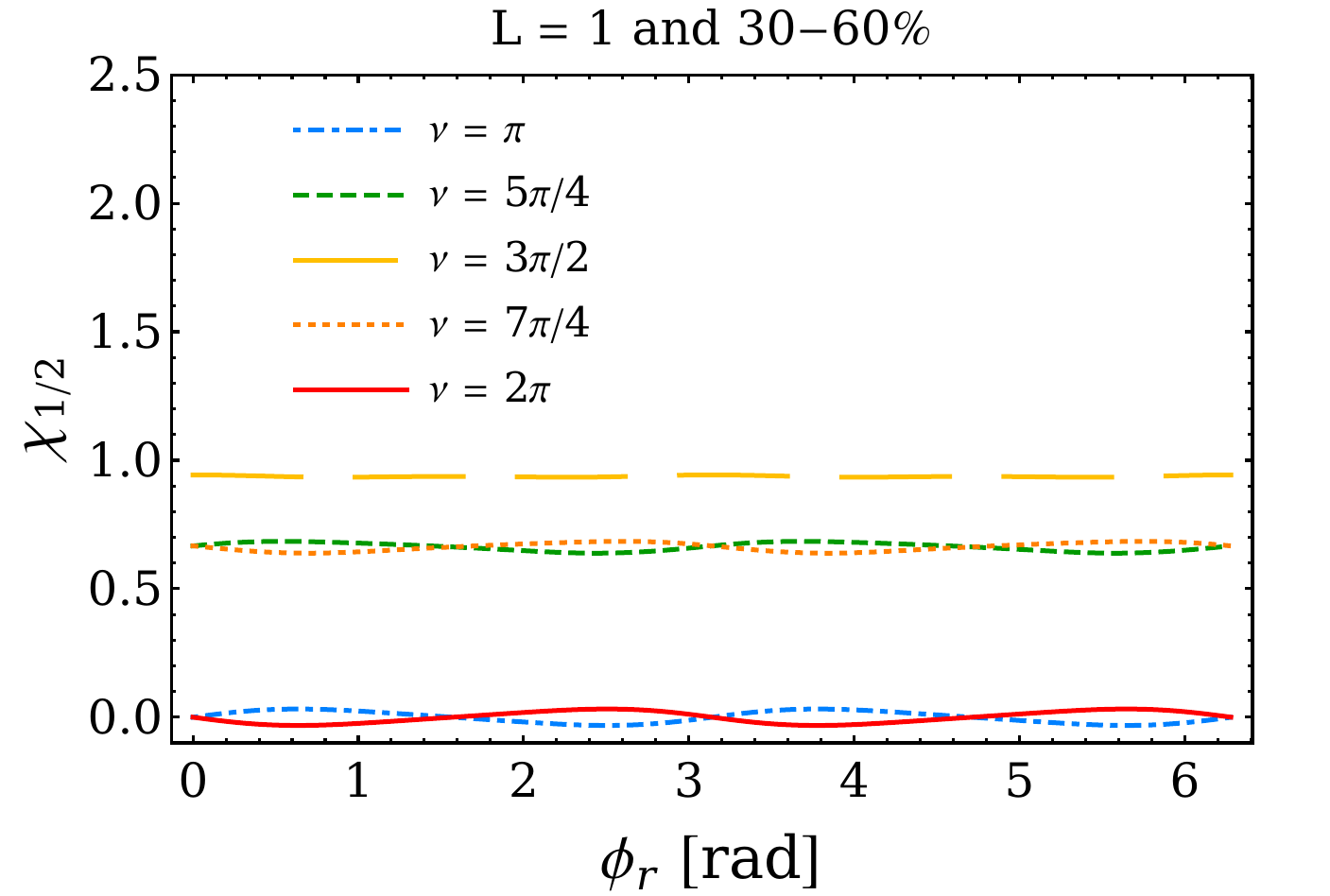}
   \caption[]{The longitudinal baryon polarization at various values of $\nu$ as a function of reaction azimuthal $\phi_r$.\label{figbarnuphi}}
\end{figure}

From equations \eqref{rhoba1} and \eqref{rhoba2}, we can observe the effects on baryon polarization for particles with spin $1/2$ under the influence of angular momentum $L=0$ and $L=1$ with $\nu_0=0$ and $\nu_0=\pi$ in Fig.\ref{figbarnuphi}.
\begin{figure}[h]
    \epsfig{width=0.49\textwidth,figure=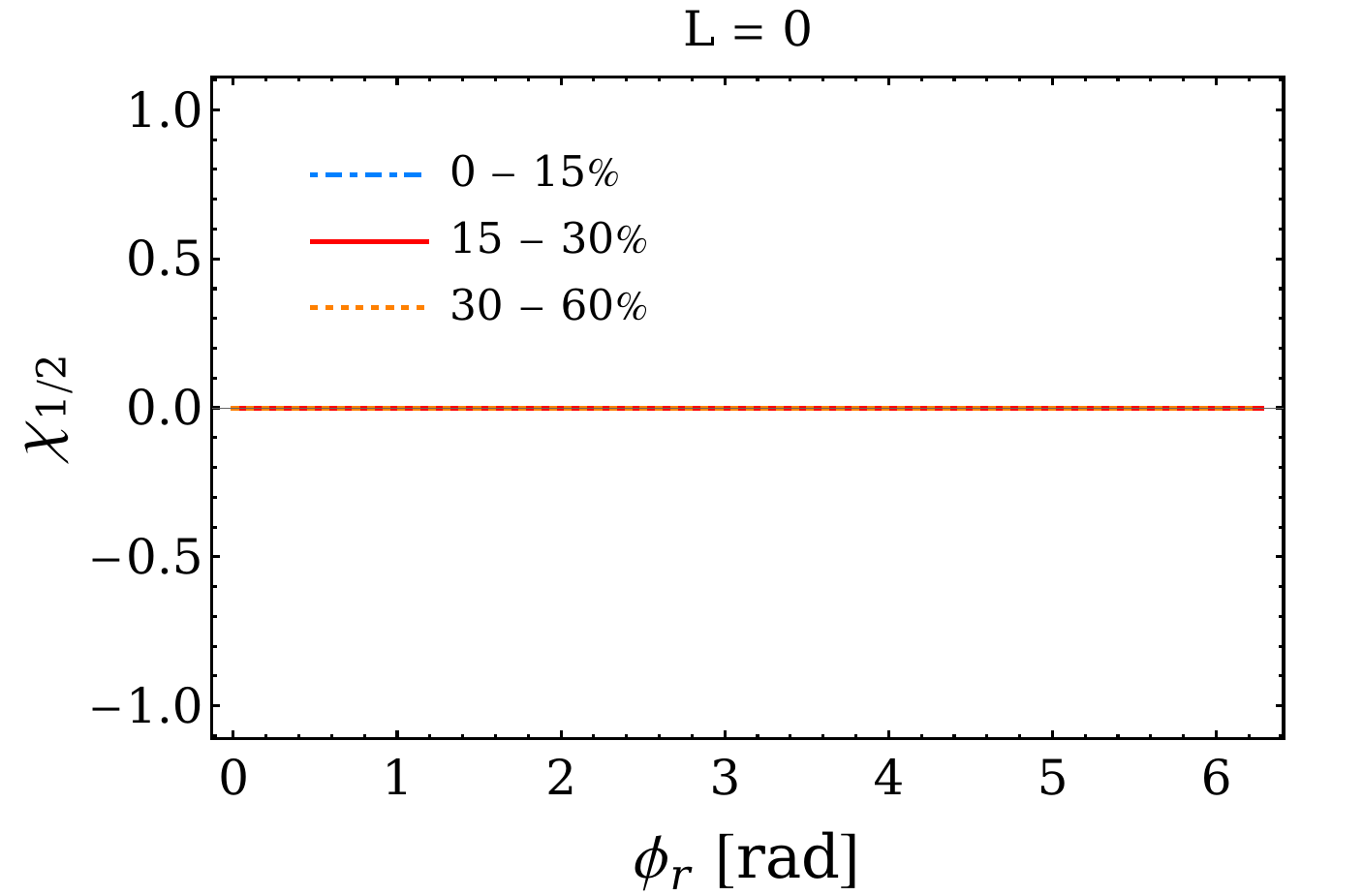}
   \epsfig{width=0.49\textwidth,figure=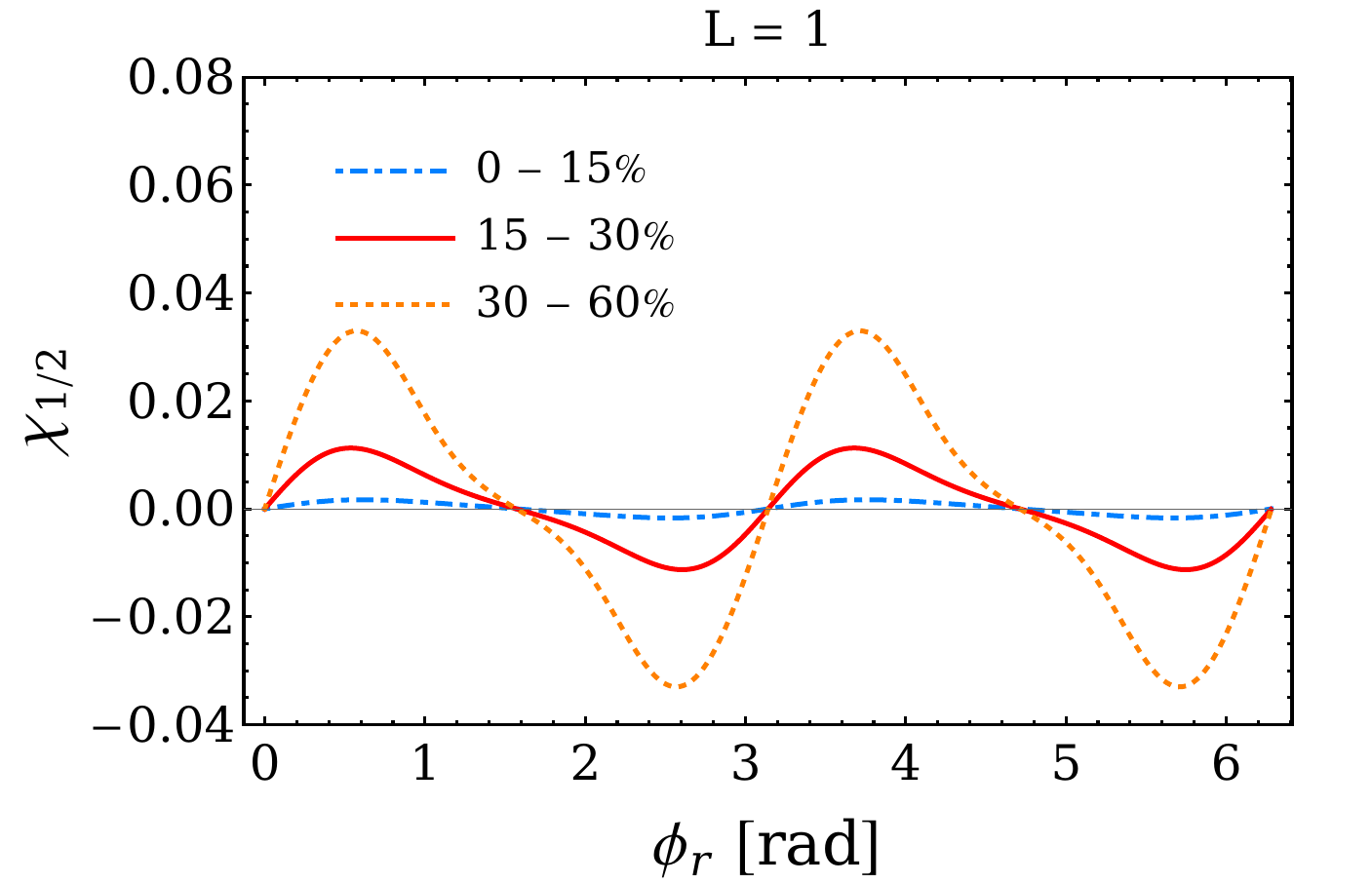}
            \caption[]{The transverse baryon polarization as a function of azimuthal angle $\phi_r$.\label{figbartrans}  }
\end{figure}

Figure \ref{figbarnuphi} one can also see that the baryon polarization oscillates with the azimuthal angle, but averages to zero, in agreement with the expectation of zero local baryon polarization.
However, Fig. \ref{figbarnuphi} further shows that the azimuthal dependence of the baryon polarization coefficient is a sensitive probe for disentangling the coefficient that regulates the transfer of angular momentum from vorticity to polarization.    In particular, we observe that varying angle between vorticity and polarization shifts the phase of the longitudinal polarization distribution with respect to the reaction plane.  This suggests that non-equilibrium between spin and vorticity could be a potential solution to the local polarization puzzle \cite{review}.  Moreover, since symmetric shear can arise from spin-vorticity non-equilibrium \cite{montediss}, this solution might be fundamentally related to the already established symmetric shear model  \cite{spinshear}

Figure \ref{figbartrans} confirms that at ultra-relativistic energies where the thermal model applies the transverse polarization is small. However, when binned in $\phi_r$ it shows considerable local variation.  This observation appears consistent with preliminary data \cite{qm2018}.

Comparing meson to baryon plots we can conclude that spin-vorticity non-equilibrium has the potential to resolve all ``puzzles`` of spin in heavy ion data.    However, our model for now is not predictive as it is non-relativistic and contains a lot of ad hoc parameters. That said, if $r_{1,0}$, $\alpha_{1,0}$, $r_{1,-1}$ and $\alpha_{1,-1}$ are found to be significant in mesons (as argued in \cite{kayman1} and calculated here in Figs \ref{r10plot},\ref{alpha10plot} ,\ref{rpmplus} and \ref{alphaplot}) Cooper-Frye formalism would imply such off-diagonal matrix elements should be zero because of Maximal Mixing, unlike coalescence), and strong evidence is found for a non-trivial modulation of $\rho_{00}$ and $\rho_{1/2,1/2}$ of mesons and baryons respectively in $\phi_r$, the calculations here will become truly predictive. This would create a direct link between experimental observables and the interaction between spin and vorticity at hadronization.  In this context, it is worth reiterating that preliminary experimental evidence \cite{qm2018} suggests that the ``zero'' baryon polarization measured at top RHIC energies may in fact conceal significant modulation, which is compatible with our rather large theoretical estimates. However, this evidence is still in the preliminary stage.

Our conclusion is that relaxing the assumption of equilibrium between vorticity and spin, well-motivated by theory \cite{causal,relax,jeon,rischke,hongo,friman}, and augmenting it with a reasonable hadronization dynamics (such as coalescence \cite{kayman1}) could have profound phenomenological consequences. This approach might help solve current ``puzzles'' in interpreting experimental data on hadronic spin, including the out-of-phase dependence of local polarization and the comparison of vector meson spin alignment with baryon polarization. 

However, this model inevitably introduces additional parameters, related specifically to spin physics (such as Wigner function $P_W$ and the vorticity projection to spin space $P_L$ in \eqref{coalmeson}). This makes detailed quantitative fits premature at this stage. Nevertheless, our work encourages further experimental work.
The most promissing observable, advocated previously, is the azimuthal dependence of spin-sensitive observables.  Specifically, significant deviations in the off diagonal elements of the spin $1$ density matrix from zero, would indicate that the Cooper-Frye freezeout dynamics is insufficient.

Another important prediction of this model is that spin $3/2$ baryons, such as $\Delta,\Omega$ will exhibit different behaviors from their spin $1/2$ counterparts.  In particular, the decay chain $\Omega \rightarrow \Lambda K \rightarrow p \pi K$, with its double weak decay, is promising. Its sequential azimuthal analysis could offer full measurement of its very large $5\times 5$ density matrix.  This will be the focus of a future project.
\section*{Acknowledgements}

 G.T. acknowledges support from Bolsa de produtividade CNPQ 305731/2023-8, Bolsa de pesquisa FAPESP 2023/06278-2. R.R.'s work was supported in part by the Polish National Science Centre Grant No. 2018/30/E/ST2/00432. K.J.G. is supported by CAPES doctoral fellowship 88887.464061/2019-00.

\end{document}